\documentclass[fleqn,usenatbib]{mnras}
\usepackage[utf8]{inputenc}
\usepackage{mathrsfs}
\usepackage{natbib}
\usepackage{amsmath}
\usepackage{amssymb}
\usepackage{graphicx}
\usepackage{hyperref}

\makeatletter

\usepackage{mathrsfs}

\newcommand{\abs}[1]{\left\vert#1\right\vert}
\newcommand{\set}[1]{\left\{#1\right\}}

\makeatother

\title[Common Envelopes and Gravitational Waves]{Gravitational waves from in-spirals of compact objects in binary common-envelope evolution}

\author[Y. B. Ginat et al.]{Yonadav Barry Ginat,$^{1}$\thanks{E-mail: ginat@campus.technion.ac.il}
Hila Glanz,$^{1}$
Hagai B. Perets,$^{1}$
Evgeni Grishin$^{1}$ and
\newauthor Vincent Desjacques$^{1}$
\\
$^{1}$Faculty of Physics, Technion -- Israel Institute of Technology,
Haifa, 3200003, Israel
}

\date{Accepted XXX. Received YYY; in original form ZZZ}

\pubyear{2019}

\begin{document}
\label{firstpage}
\pagerange{\pageref{firstpage}--\pageref{lastpage}}
\maketitle

\begin{abstract}
Detection of gravitational-wave (GW) sources enables the characterisation
of binary compact objects and of their in-spiral. However,
other dissipative processes can affect the in-spiral. Here we show
that the in-spiral of compact objects through a gaseous common-envelope
(CE) arising from an evolved stellar companion produces a novel type
of GW-sources, whose evolution is dominated by the dissipative gas
dynamical friction effects from the CE, rather than the GW-emission
itself. The evolution and properties of the GW-signals differ from those of isolated gas-poor mergers significantly. We find characteristic
strains of $\sim10^{-23}$-$10^{-21}$ ($10{\rm kpc}/{D}$) for such
sources -- observable by next-generation space-based GW-detectors.
The evolution of the GW-signal can serve as a probe of the interior regions
of the evolved star, and the final stages of CE-evolution, otherwise
inaccessible through other observational means. Moreover, such
CE-mergers are frequently followed by observable explosive electromagnetic
counterparts and/or the formation of exotic stars.
\end{abstract}


\section{Introduction}
Two gravitating bodies emit gravitational
waves (GWs) when orbiting each other, whose amplitude and frequency depend on their relative
acceleration, which, in turn, depends on their masses
and separation. Generally, larger masses and smaller separations induce stronger accelerations and stronger GW emission.
Current and upcoming GW-detectors cannot detect mergers of stellar
binaries, whose radii -- and separation at merger -- are far
below the amplitude sensitivity of the detectors at the relevant frequencies. Indeed, only sufficiently compact (and massive) stars
could serve as detectable GW sources; current studies typically focus
on the mergers of white dwarfs (WDs) \citep{LISA2017}, neutron stars
(NSs), and black holes (BHs) \citep{LIGOVIRGO2016,LIGOVIRGO2018}.
However, the stellar cores of evolved stars could serve as a novel
type of compact object (CO) to produce GW sources, once they
merge with another compact object. Such mergers can occur following
a binary common-envelope (CE) evolution of evolved stars with compact companions, leading to the in-spiral of the companion and to its
merger with the core of the evolved star \citep{Ivanova2013}.

\begin{figure}
\includegraphics[width=0.23\textwidth]{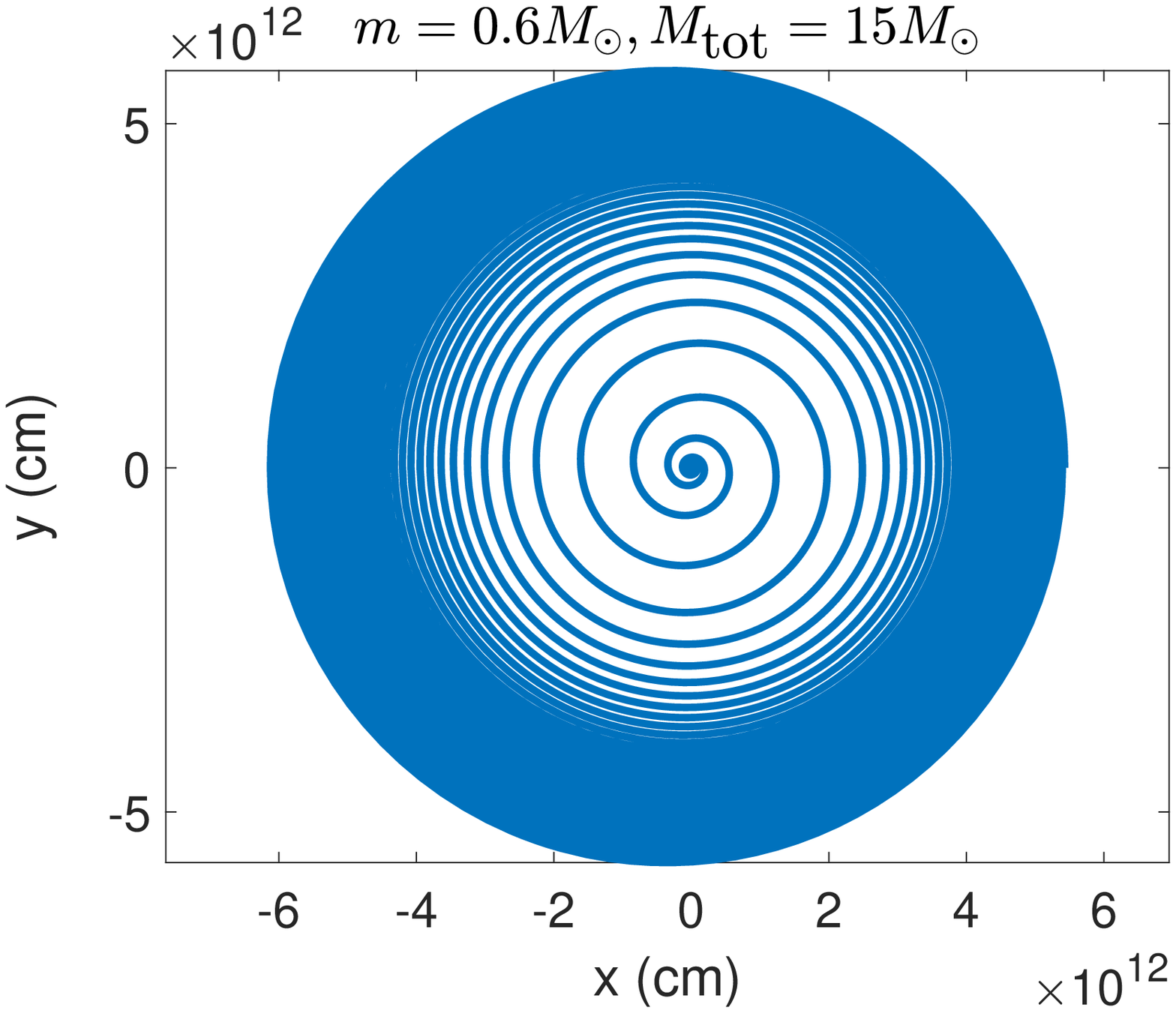} \includegraphics[width=0.23\textwidth]{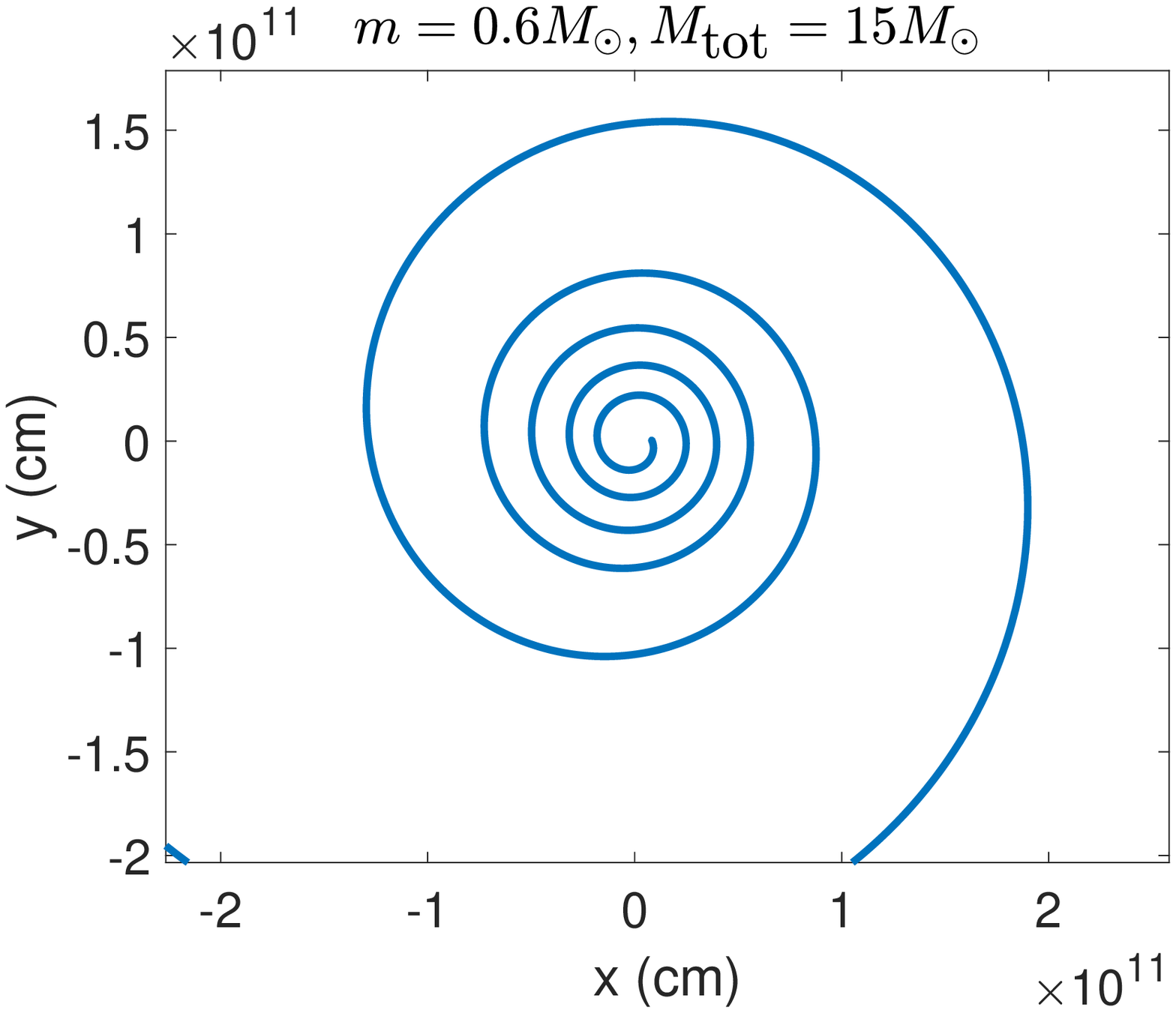}

\includegraphics[width=0.23\textwidth]{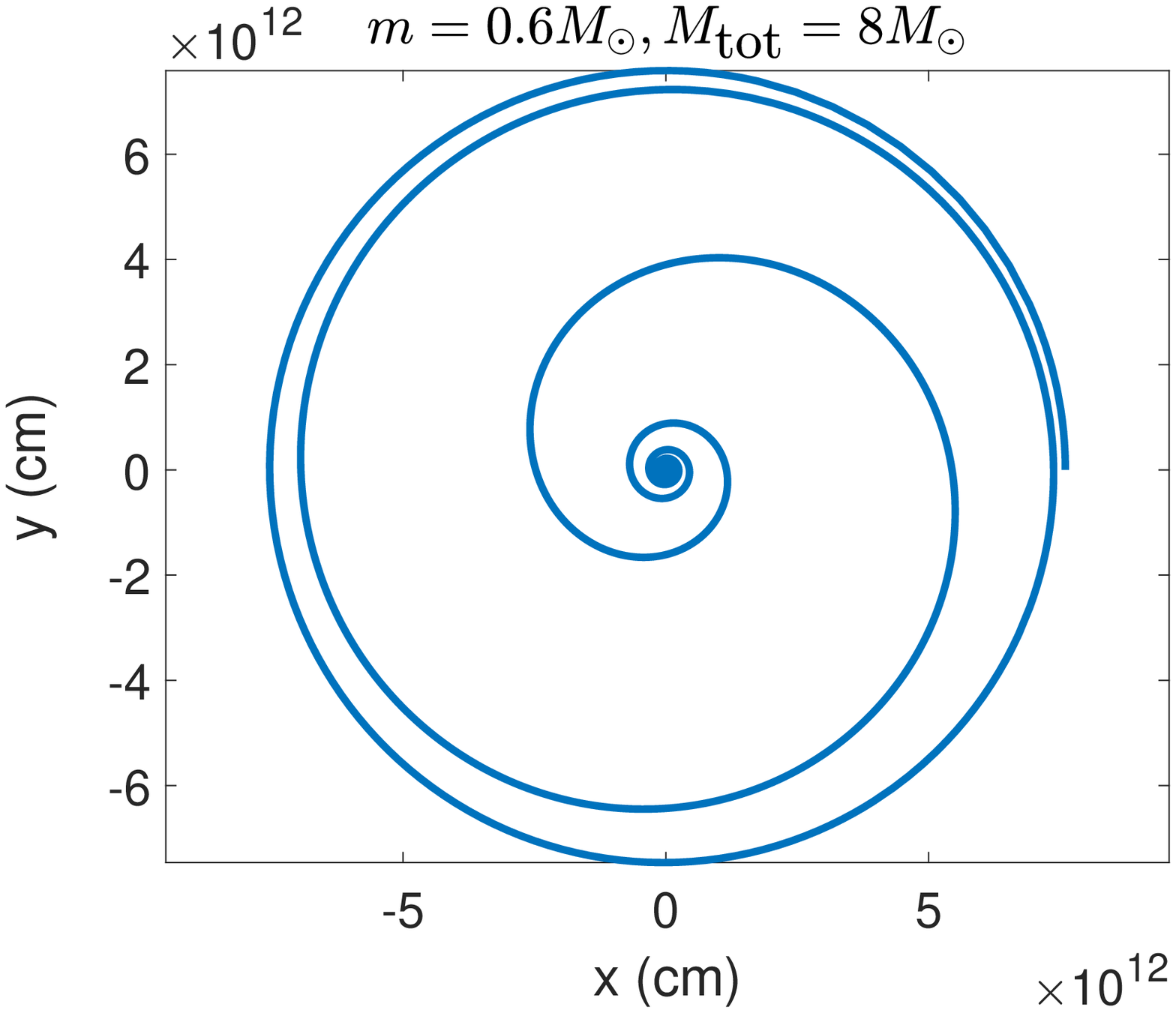} \includegraphics[width=0.23\textwidth]{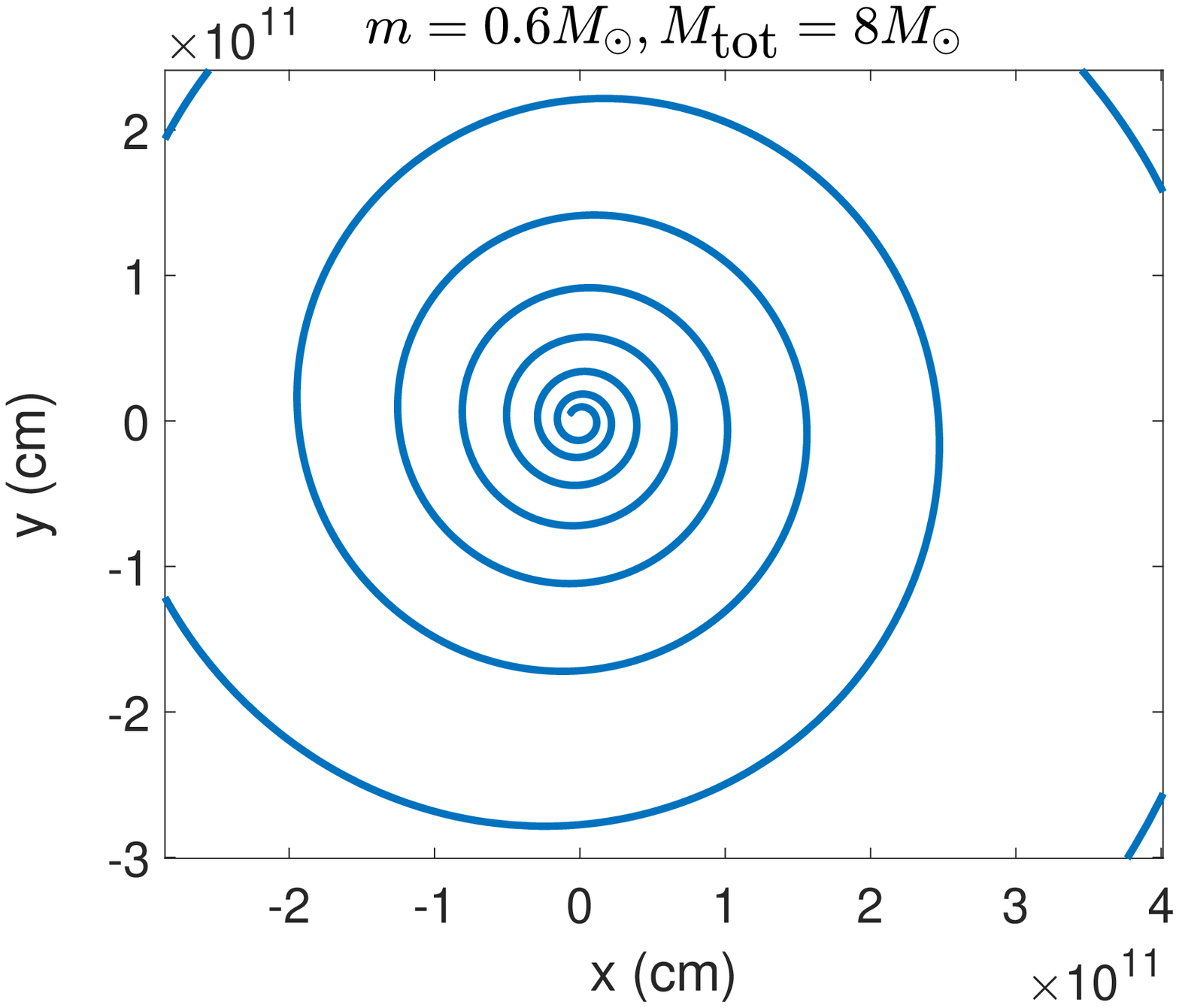}

\includegraphics[width=0.23\textwidth]{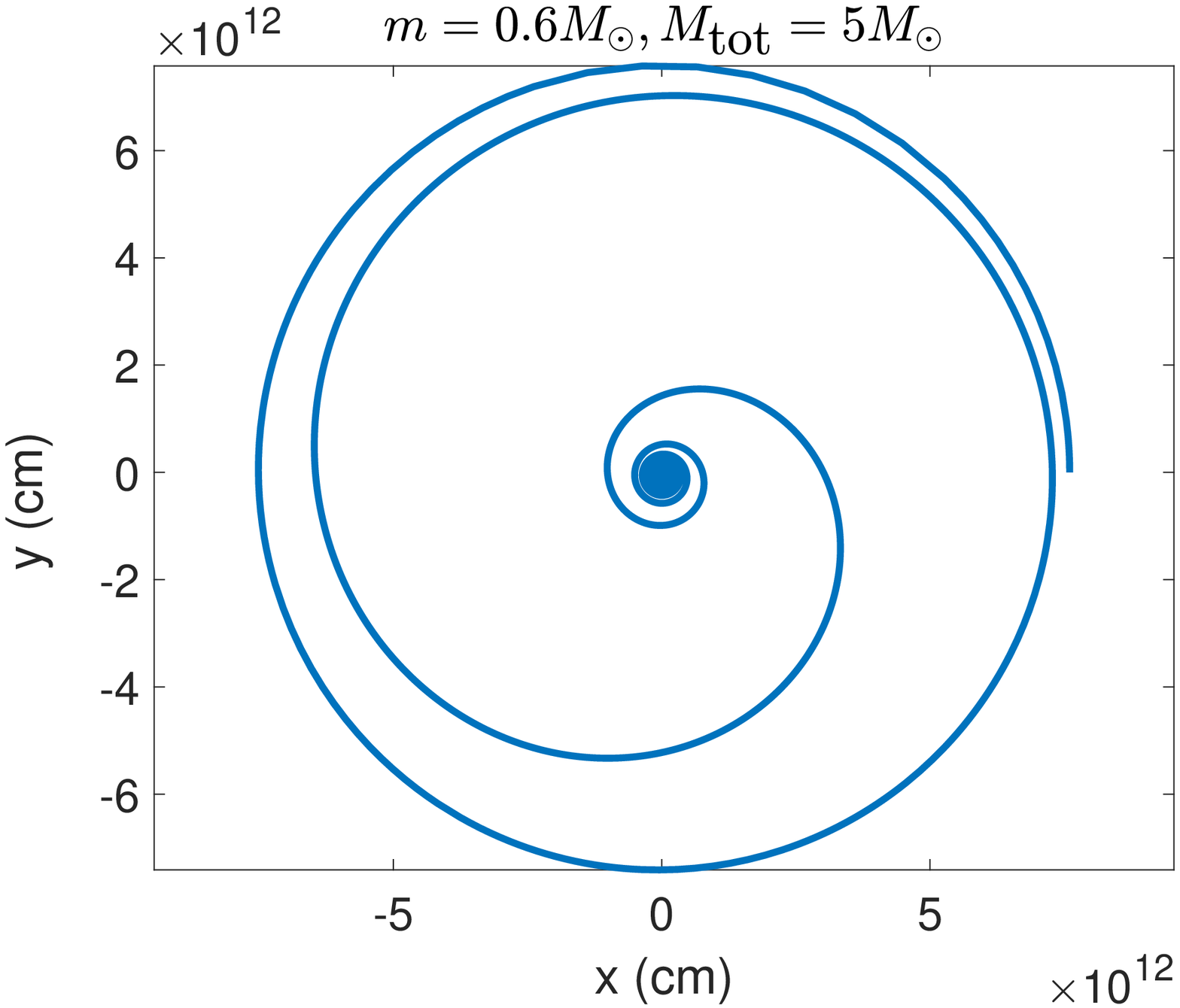} \includegraphics[width=0.23\textwidth]{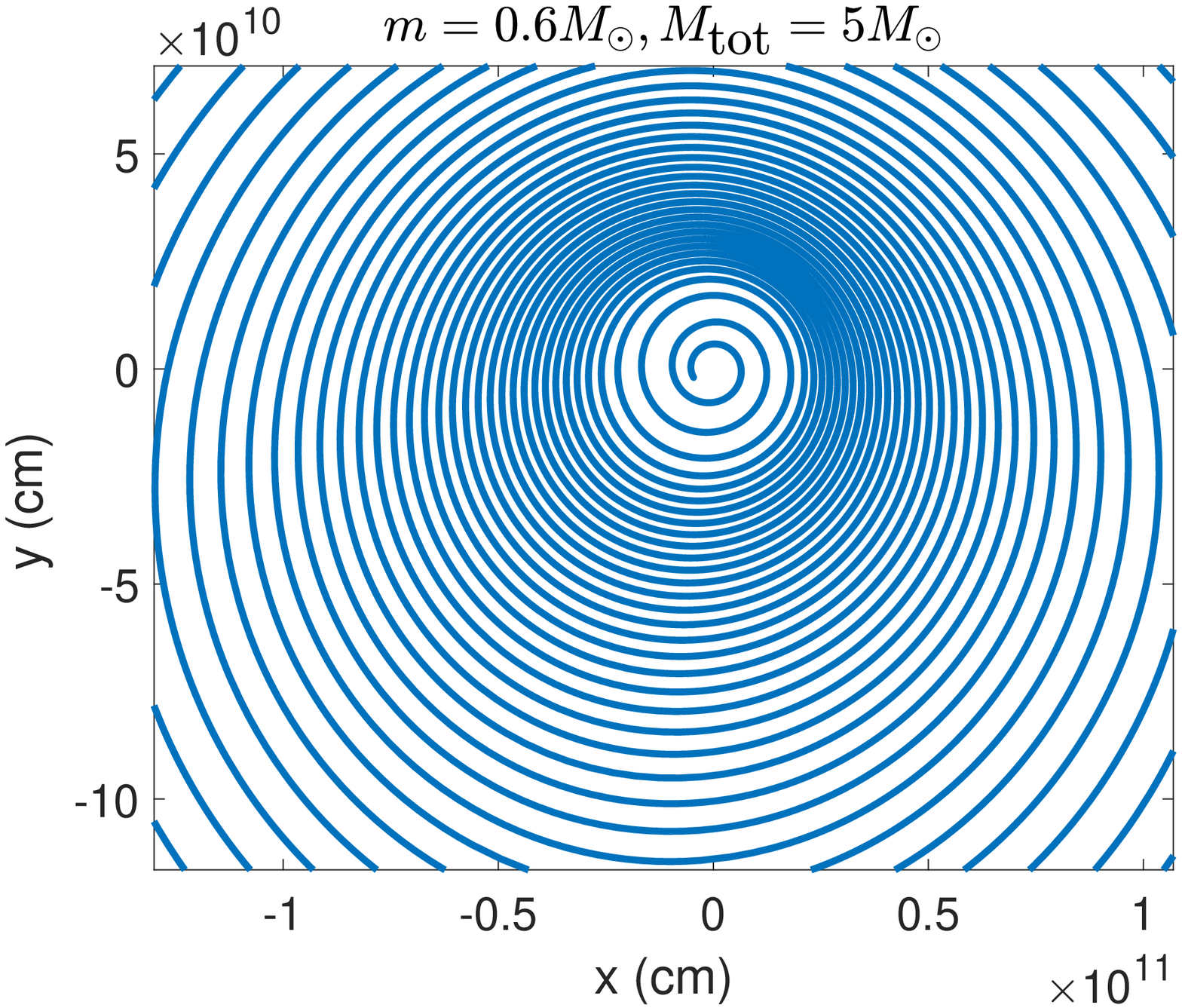}

\includegraphics[width=0.23\textwidth]{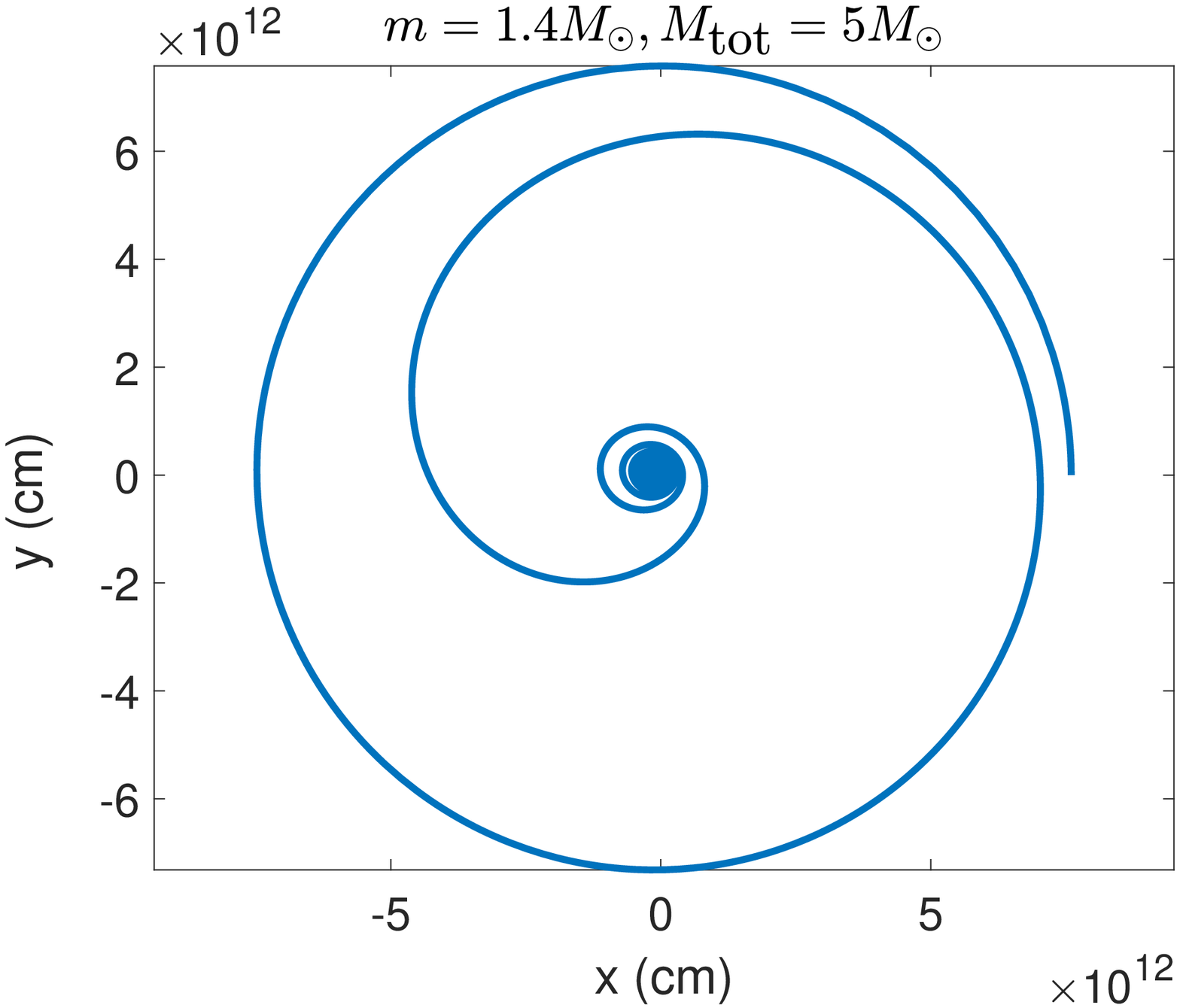} \includegraphics[width=0.23\textwidth]{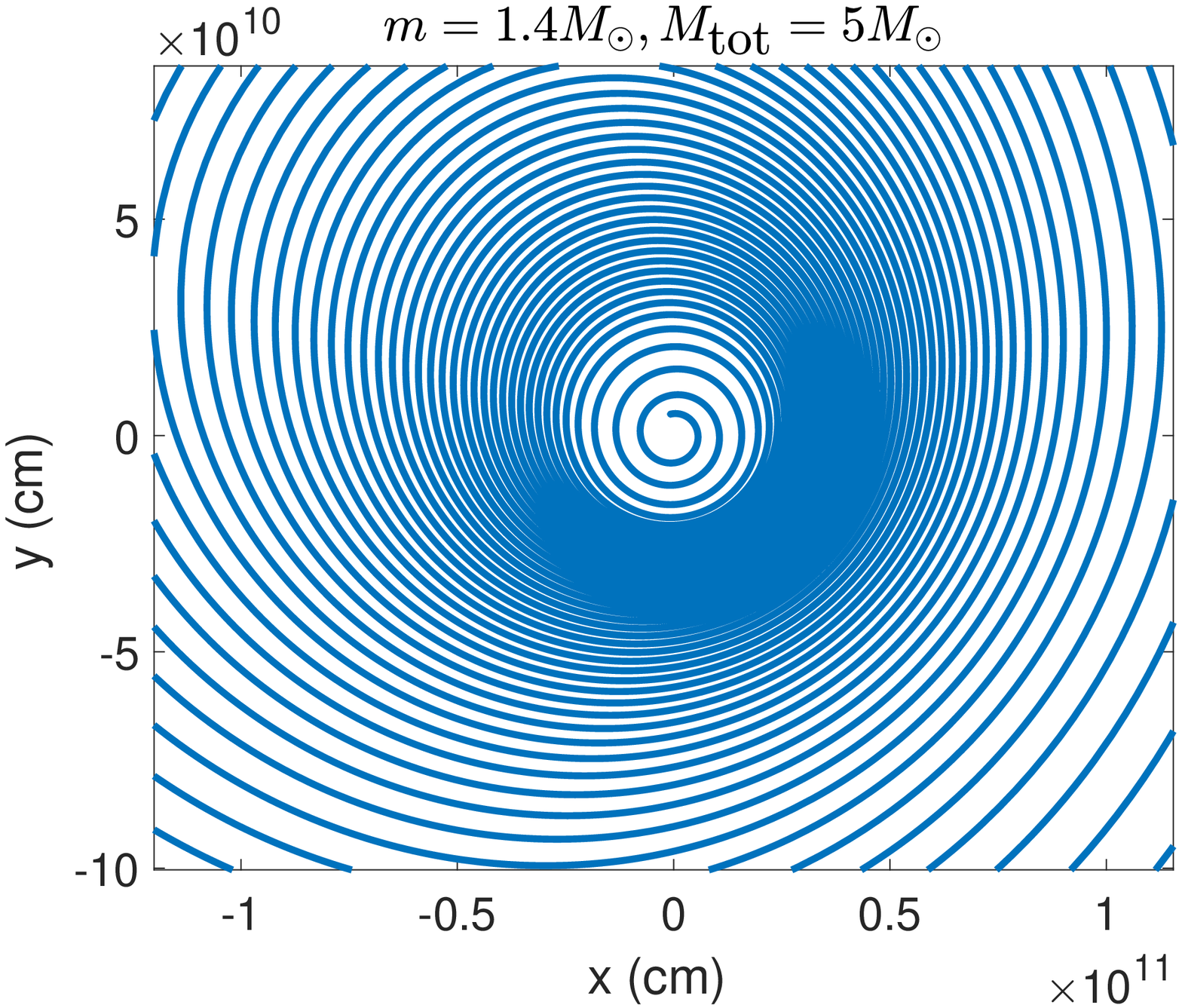}

\caption{\label{fig:orbits} The separation between the companion and the core,
for the some of the different mass combinations described in the text.
Each combination is displayed twice, with the second panel focussing
on the central region.}
\end{figure}

During the evolution of stars beyond their main-sequence evolutionary
stage, they fuse the hydrogen in their cores into helium and heavier
elements, that accumulate in compact cores. The stellar envelopes
of the stars then expand to large radii -- tens, or even up to hundreds of
solar radii (during the red-giant branch, and asymptotic giant branch,
phases of their evolution, as well as the Wolf-Rayet stage for very
massive stars). When such an evolved star has a close stellar companion
the expanded envelope engulfs the companion and, under appropriate
conditions, creates a shared envelope of gas, covering both the core of the
evolved star and the companion \citep{Ivanova2013}. The stellar companion
is then thought to spiral inside the envelope due to its gravitational
interaction with the gas (see figure \ref{fig:orbits} for depictions of such in-spirals). The CE phase
could result in either an ejection of the gaseous envelope before
the companion arrives at the core, or in a final merging
of the companion with the compact core. The former occurs
when the evolved stellar envelope is not sufficiently massive and
dense compared with the stellar companion, and \emph{vice versa}.
Here we show that when the stellar companion is compact, and the core is also sufficiently compact, CE
evolution leads to various possible CO-core mergers with
detectable GW signals (see \citealt{NazinPostnov1995,NazinPostnov1997}
for an initial study for neutron star companions). Note that significant fractions of the envelope may be ejected on much longer non-dynamical timescales \citep{Gla+18,Mic+19}; this, however is always preceded by the dynamical phase.

The in-spiral of a compact object proceeds on a dynamical
timescale when the dynamical friction due to the CE is sufficiently
strong. The closest approach accessible for a CO-core binary
occurs when the CO reaches the tidal radius of
the core $r_{t}\sim\left(M_{\textrm{CO}}/M_{\textrm{core}}\right)^{1/3}{\rm R_{\textrm{core}}}$,
at which point the latter is disrupted tidally. To characterise the GW-detectability
of such binary mergers, one needs to derive the strain and frequency
of the GW emission of such binaries close to the final radius. One also needs to account for the evolution due to the interaction
with the gaseous envelope. This process is special, in-so-far as the
driving force behind the in-spiral is not energy-loss to gravitational
waves, but rather interaction with an environment. As such the gravitational-wave
signal should reflect the properties of the environment, and allow
for potentially direct observations of
such otherwise-inaccessible environments/processes \citep{Pani2015,Fedrowetal2017}.

To study such GW sources we model
the evolution as two point masses orbiting inside an envelope with
a given constant density profile, and we neglect the back-reaction
of the motion of the system on the envelope itself. We model
the in-spiral through the effects of gas dynamical friction (GDF) using an approach originally presented by \cite{Ostriker1999}, rather than the approach taken previously
by \citet{NazinPostnov1995}, and include 2.5 post-Newtonian corrections
to the gravitational interaction of the CO and the core. The resultant GW signal would depend
on the density profile of the CE, thereby enabling the utilization of gravitational waves to probe the properties of the interior of the red-giant/CE.

The in-spiral is driven by GDF and not by aerodynamic drag, due to the
high ratio between the mass of the in-spiralling CO and its geometric
cross-section. \citet{GP15} found that the critical
transition radius between GDF-dominated evolution and aerodynamic-drag
dominated evolution is $R_{c}\propto v_{{\rm rel}}/\sqrt{G\rho_{m}}$,
where $v_{{\rm rel}}$ is the relative velocity between the CO and
the gas, and $\rho_{m}$ is the density of the CO. The proportionality
constant depends on the dimensionless Reynolds and Mach numbers of
the flow, and is of order unity for most plausible situations.

In our case, we take an initial relative velocity of $\sim100~\textrm{km s}^{-1}$
(similar to the Keplerian velocity around a massive star at $1\ {\rm AU}$).
For a typical WD density of $\rho_{{\rm WD}}\approx10^{6}\ {\rm g\ cm^{-3}}$,
we get a critical radius of a few percent of the WD radius,
$R_{c}\approx0.05R_{{\rm WD}}$, with $R_{{\rm WD}}=0.01R_{\odot}$.
For typical NS densities of $\rho_{{\rm NS}}=10^{14}\ {\rm g\ cm^{-3}}$,
the critical radius is around $4\times10^{3}~\textrm{cm}$, much less
than a NS radius. Therefore, for compact objects embedded in envelopes of giant stars, GDF dominates over
aerodynamic drag.

We begin by describing the set-up
of our models and the equations of motion, which we numerically integrate
to calculate the GW signatures of the CE gravitational-wave
(CEGW) sources, exploring various configurations of CE-binaries and
COs. We then discuss the properties of such CEGW sources
and the possibility of detecting them, and then we summarise.

\section{Model Description}
\label{sec:set-up}
We work in the centre-of-mass frame of the binary, and neglect any back-reaction of the companion
on the envelope -- we take the density of the envelope
to be constant in time, and `glued' to the core of the star, which
we model as a point mass \footnote{This approximation is more consistent for cases where the envelope
is significantly more massive than the companion (see below). In practice, the
CE evolution changes the structure of the envelope even when the envelope
is more massive.}. This system may be described as an effective one body, whose position
$\mathbf{r}$ describes the relative separation between the companion
and the core. Its equations of motion are
\begin{equation}
\ddot{\mathbf{r}}=-\frac{G(M+m)}{r^{3}}\mathbf{r}-\frac{GM_{\textrm{env}}(r)}{r^{3}}\mathbf{r}-F(\mathbf{r},\mathbf{v})\mathbf{v}+\mbox{P.N.},\label{eqn:motion}
\end{equation}
where $M_{\textrm{env}}(r)$ is the mass of the envelope inside a
sphere of radius $r$ (excluding the core), $M=M_{\textrm{core}}$
is the core mass, and $m=M_{\textrm{CO}}$ is the companion mass,
and `P.N.' denotes post-Newtonian terms. We further define $M_{\textrm{tot}}=M+M_{\textrm{env}}(R)$,
where $R$ is the radius of the envelope. $F$ describes the effects
due to GDF, which model like \citet{Ostriker1999}:
\begin{equation}
F(\mathbf{r},\mathbf{v})=\frac{2\pi G^{2}m\rho(\mathbf{r})}{v^{3}}\begin{cases}
\ln\left(\frac{1+\mathscr{M}}{1-\mathscr{M}}e^{-2\mathscr{M}}\right), & \mathscr{M}<1\\
\ln\left(\Lambda^{2}-\frac{\Lambda^{2}}{\mathscr{M}^{2}}\right), & \mathscr{M}>1.
\end{cases},
\end{equation}
where $\mathscr{M}=v/c_{s}$ is the Mach number, and $c_{s}$ is
the local sound-speed of the envelope. The Coulomb logarithm $\ln\Lambda$
is given by $\Lambda=b_{\max}/b_{\min}$ \cite[p. 835]{Binney},
where $b_{\min}=\max\set{Gm/v^{2},r_{\textrm{coll}}}$; $r_{\textrm{coll}}$
is the radius at which the compact object collides with the core. We take $b_{\max} = 2r$, as in \citet{KimKim2007} rather than $b_{\max} = R_{\textrm{env}}$, because the density outside $r$ is a lot smaller than the density inside $r$.

\subsection{Time-Scales}
During the most of the in-spiral $v>c_{s}$, and the dynamical friction force is well approximated
by
\begin{equation}
F(\mathbf{r},\mathbf{v})\sim N\frac{G^{2}m^{2}\rho(r)}{v^{3}},\label{eqn:dynamical fricntion}
\end{equation}
where $N$ is a dimensionless constant, proportional to the Coulomb
logarithm $\ln\Lambda$. Let us try to get an order-of-magnitude estimate
on the decay time: as $F_{DF}=-Fv\sim v^{-2}$, we can treat the orbit
as approximately circular. Taking $v^{2}\sim G(M+m+M_{\textrm{env}}(r))/r$, gives a collapse time-scale of \begin{equation}
\tau_{DF}=\frac{\mu v^{3}}{NG^{2}m^{2}\bar{\rho}}\sim\frac{\mu(M+m+M_{\textrm{env}}(r))^{3/2}}{Nr^{3/2}\sqrt{G}m^{2}\bar{\rho}}.
\end{equation}
If we take $\rho=\bar{\rho}$, we need to take $r$ as the radius
at which the density is equal to the average density, which we denote
by $r=CR$, where $C<1$ describes the concentration of the system.
Setting $M_{\textrm{env}}(r)\approx M_{\textrm{env}}$, we have
\begin{equation}
\begin{aligned}
\tau_{DF} & =\frac{4\pi\mu(M+m+M_{\textrm{env}})^{3/2}a^{3/2}}{3C^{3/2}N\sqrt{G}m^{2}M_{\textrm{env}}}\\
 & =0.1~\textrm{year}\left(\frac{M+m+M_{\textrm{env}}}{2M_{\odot}}\right)^{3/2}\left(\frac{R}{100R_{\odot}}\right)^{3/2} \\ & \times \left(\frac{M_{\odot}}{m}\right)^{2}\left(\frac{M_{\odot}}{M_{\textrm{env}}}\right)\left(\frac{0.1}{C}\right)^{3/2}\left(\frac{4\ln10}{N}\right).
\end{aligned}
\end{equation}

\subsection{Wave-Forms}
During the in-spiral the binary emits gravitational waves due
to a changing quadrupole moment. As $r$ decreases, the orbital
acceleration increases, and so does the GW-amplitude, right up to
the final merger. The leading-order GW-strain $h_{ab}(t)$ is given by the quadrupole
formula:
\begin{equation}
h_{ab}^{TT}(t)=\frac{2G}{Dc^{4}}\ddot{Q}_{ab}^{TT}(t_{\textrm{ret}}),\label{eqn:quadrupole}
\end{equation}
where $TT$ denotes the transverse-traceless gauge, and an upper $TT$
on $Q$ implies a projection on the direction of observation; $Q_{ab}=M^{ab}-\delta_{ab}M^{kk}/3$
is the quadrupole moment, and $t_{\textrm{ret}}=t-D/c$ is the retarded
time, with $D$ being source-to-detector distance.
Even though one of the bodies is not point-like, it is spherical,
so the mass moment is
\begin{equation}
M^{ij}=mx_{1}^{i}(t)x_{1}^{j}(t)+M_{\textrm{tot}}x_{2}^{i}(t)x_{2}^{j}(t)+\frac{1}{3}\delta^{ij}C,
\end{equation}
where $C$ is a higher moment of the density $\rho$, and $\mathbf{x}_{1,2}$
is the position of the centre of the CO/giant. The second term does not contribute to $Q_{ab}$, so we can ignore it. The contribution
of the evolved star is thus equal to that of a point mass, and we may therefore
use the formula for $M^{ij}$ of a binary system in the centre-of-mass frame (e.g. given by equation (3.72) of \citealt{Maggiore}). The
solution to equation \eqref{eqn:motion} defines the wave-form, so
the problem of calculating the wave-form reduces to solving the
equations of motion.

\section{Numerical Calculations}
\label{sec:numerics}
To calculate the CEGW signatures, we integrate equation \eqref{eqn:motion}
numerically using a Runge-Kutta integrator (MATLAB's ode113). We added relativistic corrections to the motion of the core
and the companion up to 2.5PN (to account for GW dissipation)\footnote{These have significant contributions only when they are extremely
close to each other, so the post-Newtonian terms do not include the
gravity of the gas in the envelope.}, as given by \citet{LincolnWill1990}. The orbit can still be described
by a point particle with mass $\mu=mM_{\textrm{tot}}/(m+M_{\textrm{tot}})$,
moving under the influence of the gravitational field a mass $M+m$ and inside the envelope's potential. Here
we considered a few combinations of $m,M$ and $M_{\textrm{env}}$
(see table \ref{tab:snr} and figure \ref{fig:orbits}).

The density profiles of the stars were obtained from stellar
evolution models, produced by the MESA code \citep{Paxtonetal2011,Paxtonetal2013,Paxtonetal2015}, evolved from initial masses
of $15$, $8$ and $5$ ${\rm M_{\odot}}$, until they reached $110R_{\odot}$; the stars retained almost all of their original mass. Examples of orbits are shown in figure \ref{fig:orbits}.

\begin{figure*}
\includegraphics[width=0.46\textwidth]{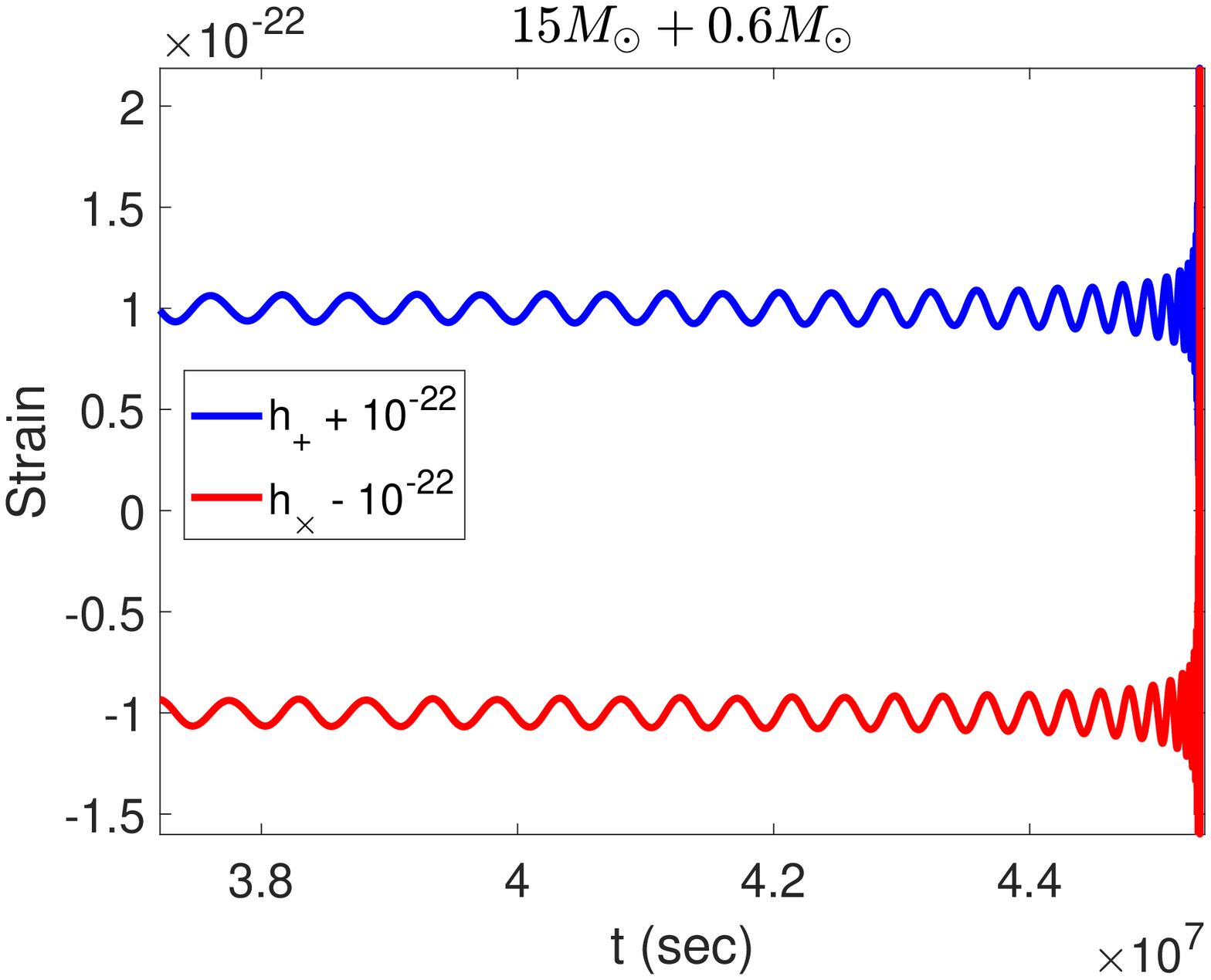} \includegraphics[width=0.46\textwidth]{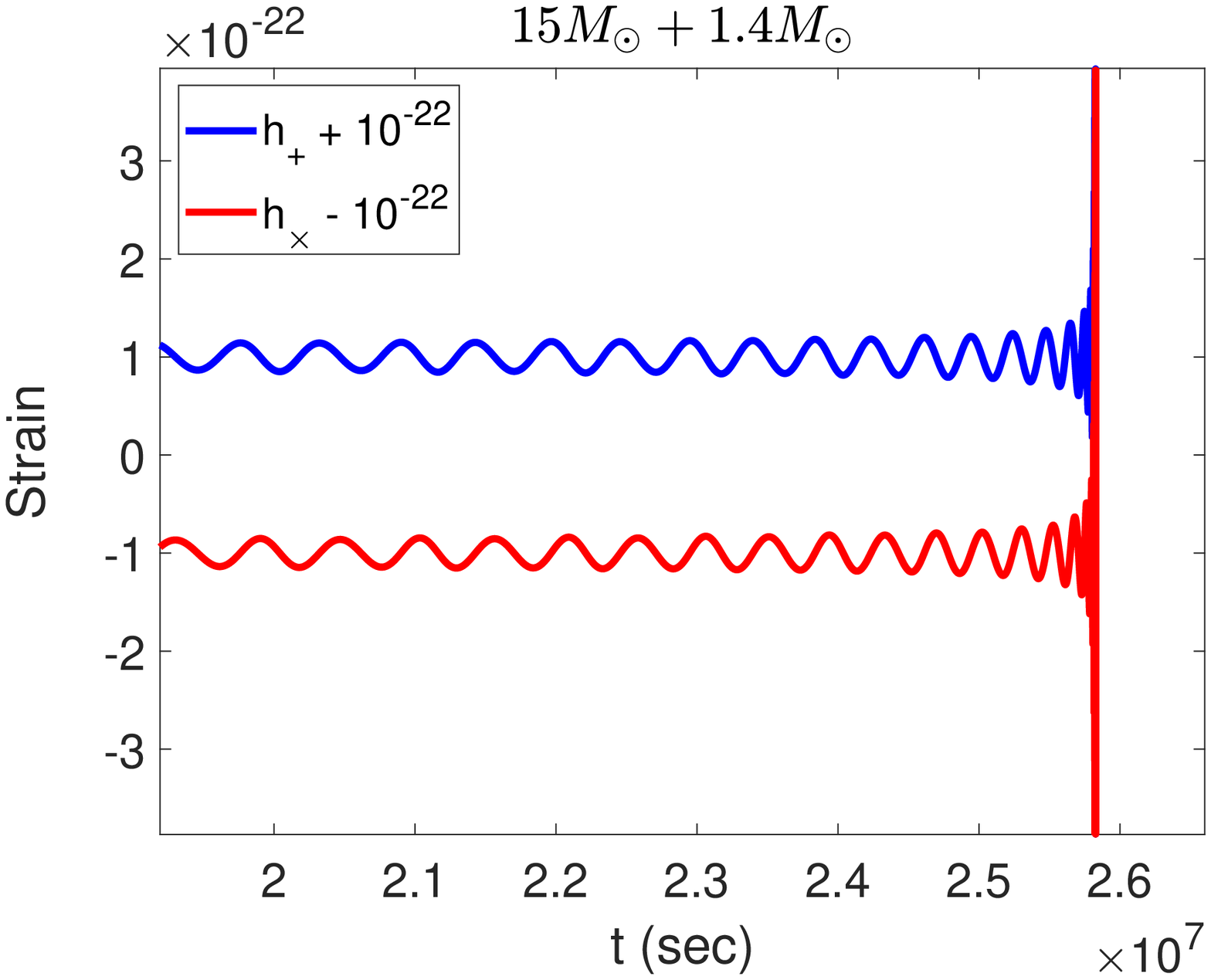}

\includegraphics[width=0.46\textwidth]{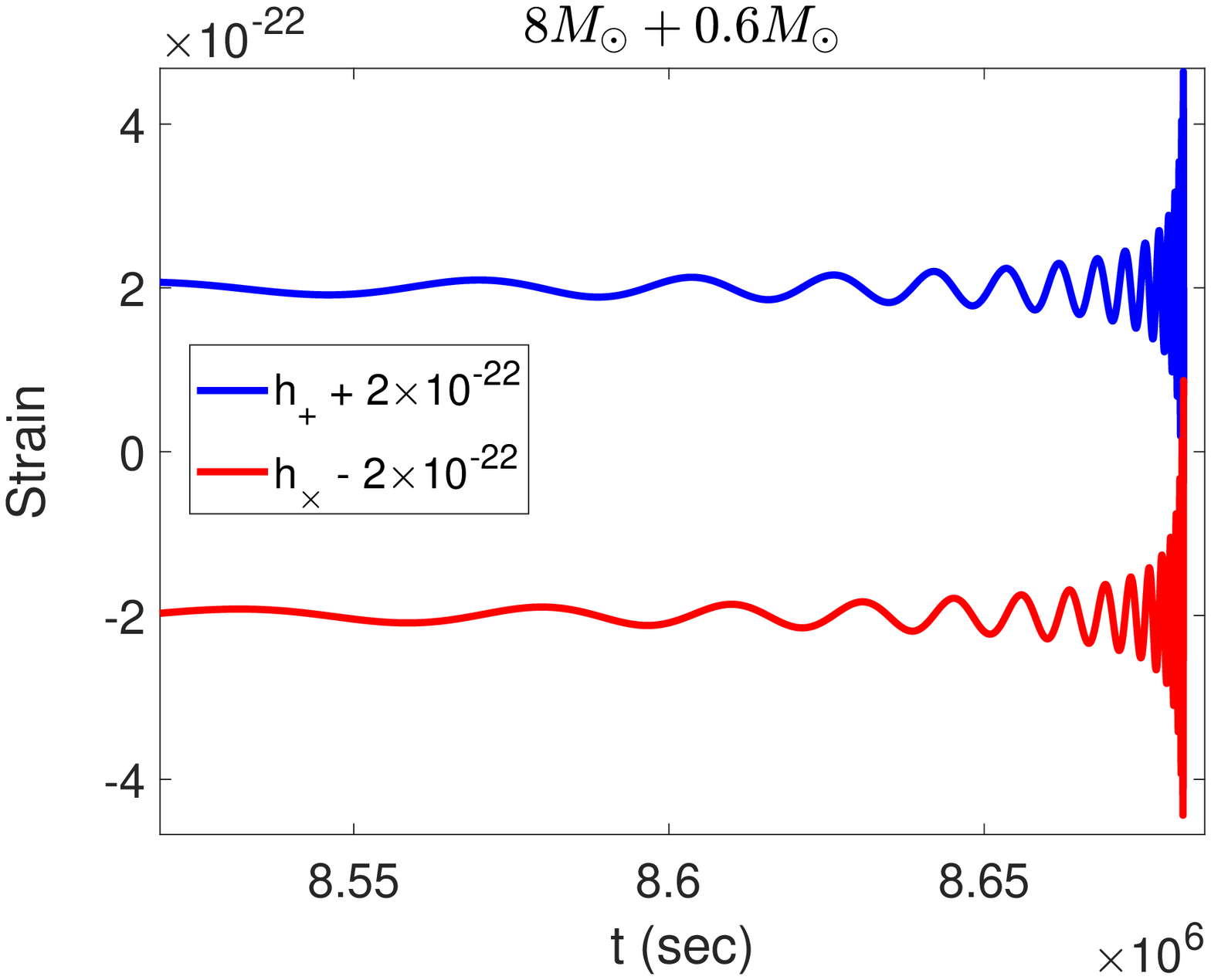} \includegraphics[width=0.46\textwidth]{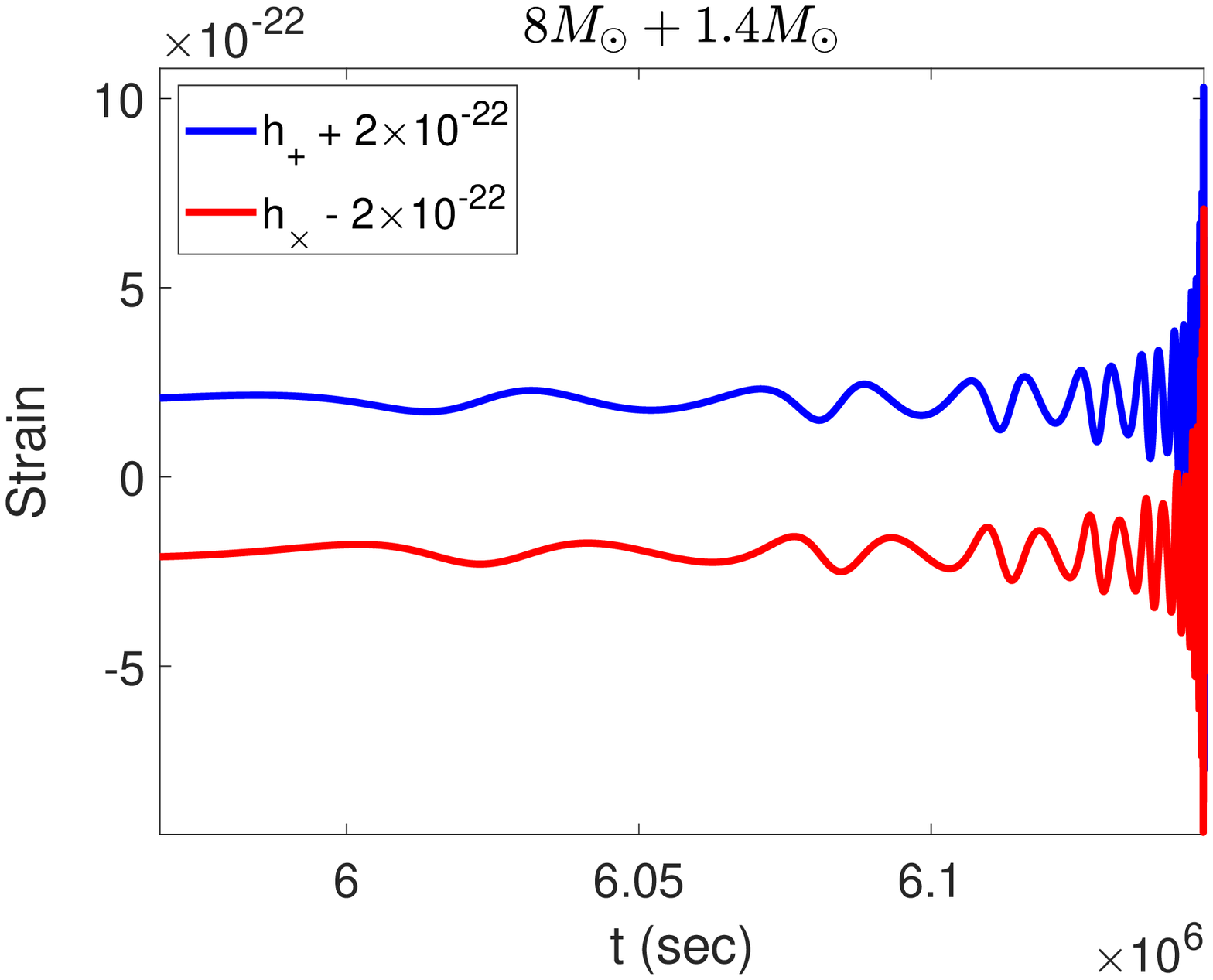}

\includegraphics[width=0.46\textwidth]{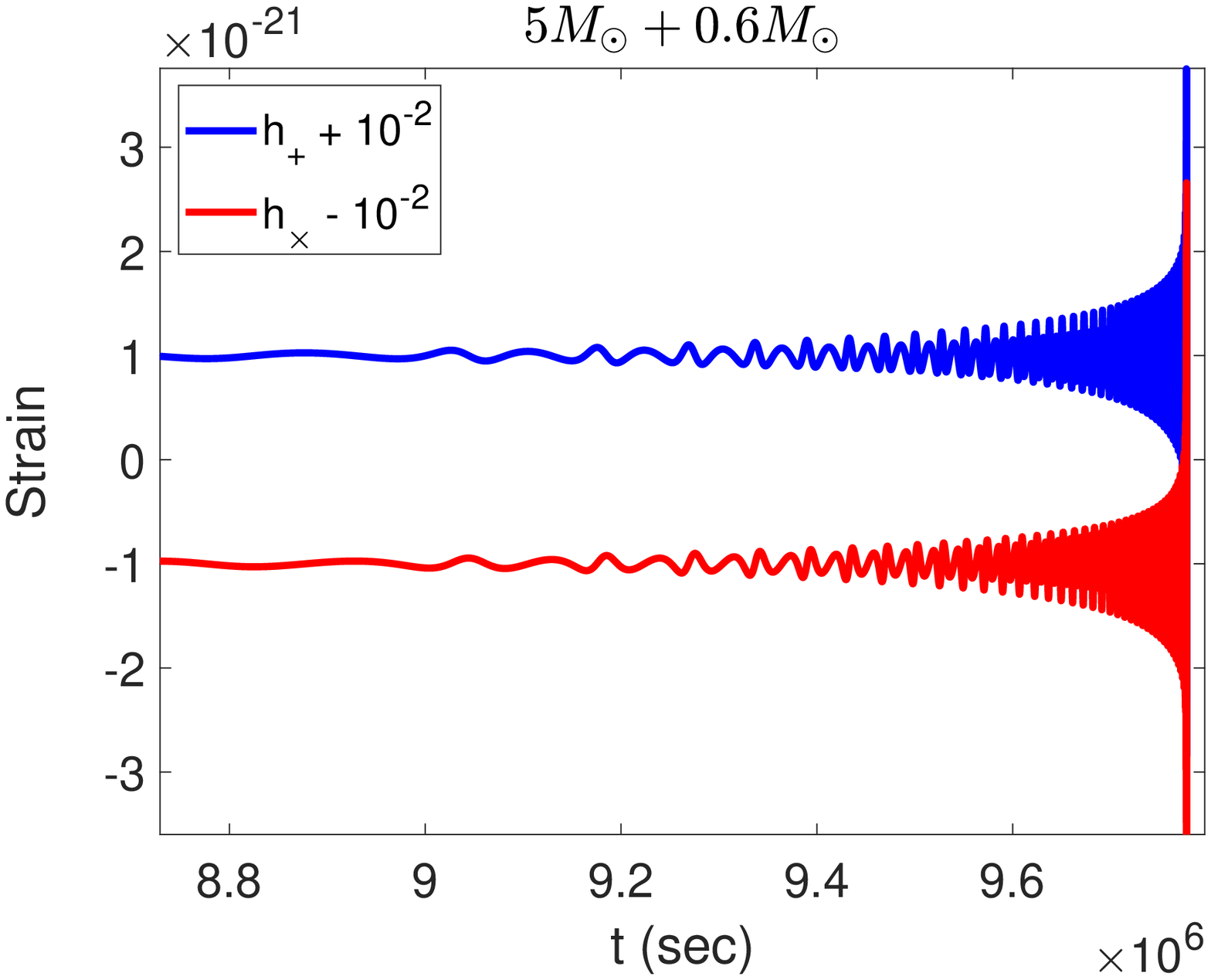} \includegraphics[width=0.46\textwidth]{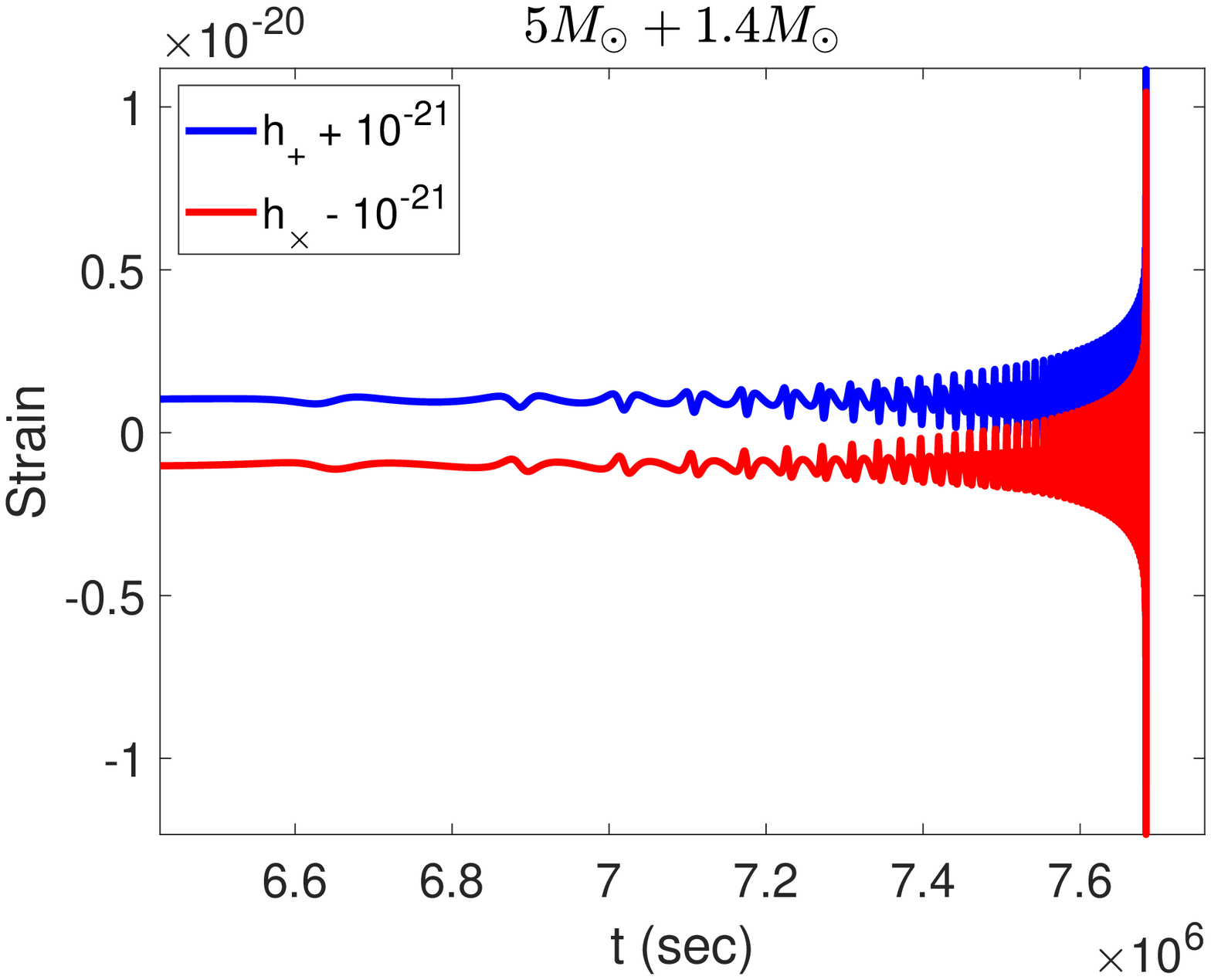}
\caption{\label{fig:wave-forms} CEGWs (both polarizations), shown for the different
combinations of COs and evolved star binaries, computed
at a distance of $10\textrm{kpc}$.}
\end{figure*}

For each of these models we calculate the emitted waves, with $m=0.6M_{\odot}$ representing a WD companion,
and $m=1.4M_{\odot}$ -- a NS. The detector is assumed to be at a distance of
$10~\textrm{kpc}$. The resulting wave-forms are shown
in figure \ref{fig:wave-forms}. All in-spirals show a characteristic
evolution beginning with regular low-amplitude oscillations at the early phases, which then gradually increase
in frequency and amplitude down to the final plunge accompanied by
a high amplitude burst.

\section{Detectability}
\label{sec:detectability}
Let us proceed to consider the detectability of such CEGWs: the size of the core
is comparable to (or larger than) that of WDs, and the orbital frequency
of the binary before its final merger is therefore lower than the
detection range of aLIGO. However, as we show below, such
GW sources are detectable by next-generation space-based GW-detectors,
such as LISA.

Whether the predicted signal is detected or not is determined by its characteristic strain, which is related to the signal-to-noise
ratio (SNR). If $S_{n}(f)$ is the noise power-spectrum density of
a detector, and $h$ is the signal (without noise) then the
SNR is given by \citep{Mooreetal2015}
\begin{equation}
\left(\frac{S}{N}\right)^{2}=4\int_{0}^{\infty}\frac{\abs{\tilde{h}(f)}^{2}}{S_{n}(f)}\mathrm{d}f=\int_{0}^{\infty}\frac{h_{c}^{2}(f)}{h_{n}^{2}(f)}\mathrm{d}f,
\end{equation}
where $\tilde{h}(f)=\int_{-\infty}^{\infty}e^{-2\pi \mathrm{i}ft}h(t)\mathrm{d}t$.
The characteristic strain is $h_{c}(f)=2f\tilde{h}(f)$; so,
if $h_{n}=\sqrt{fS_{n}(f)}$ signifies the sensitivity of the detector,
the area between the characteristic strain and the sensitivity curve
in logarithmic scale is the SNR. These are calculated using a discrete
fast Fourier transform, applying a Tukey window \citep{LIGOVirgo2016b} to alleviate spectral leakage.

\begin{figure*}
\includegraphics[width=0.46\textwidth]{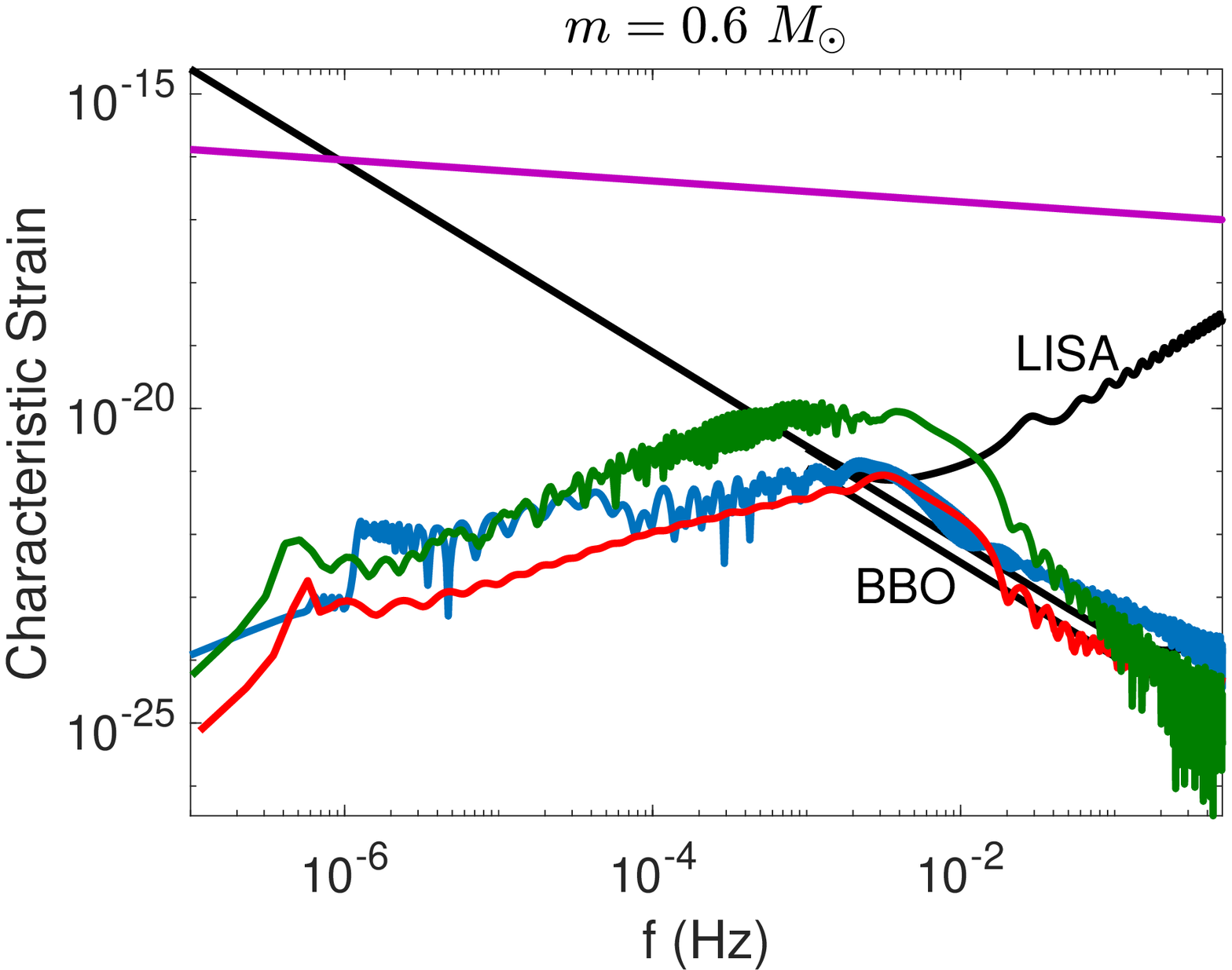}
\includegraphics[width=0.46\textwidth]{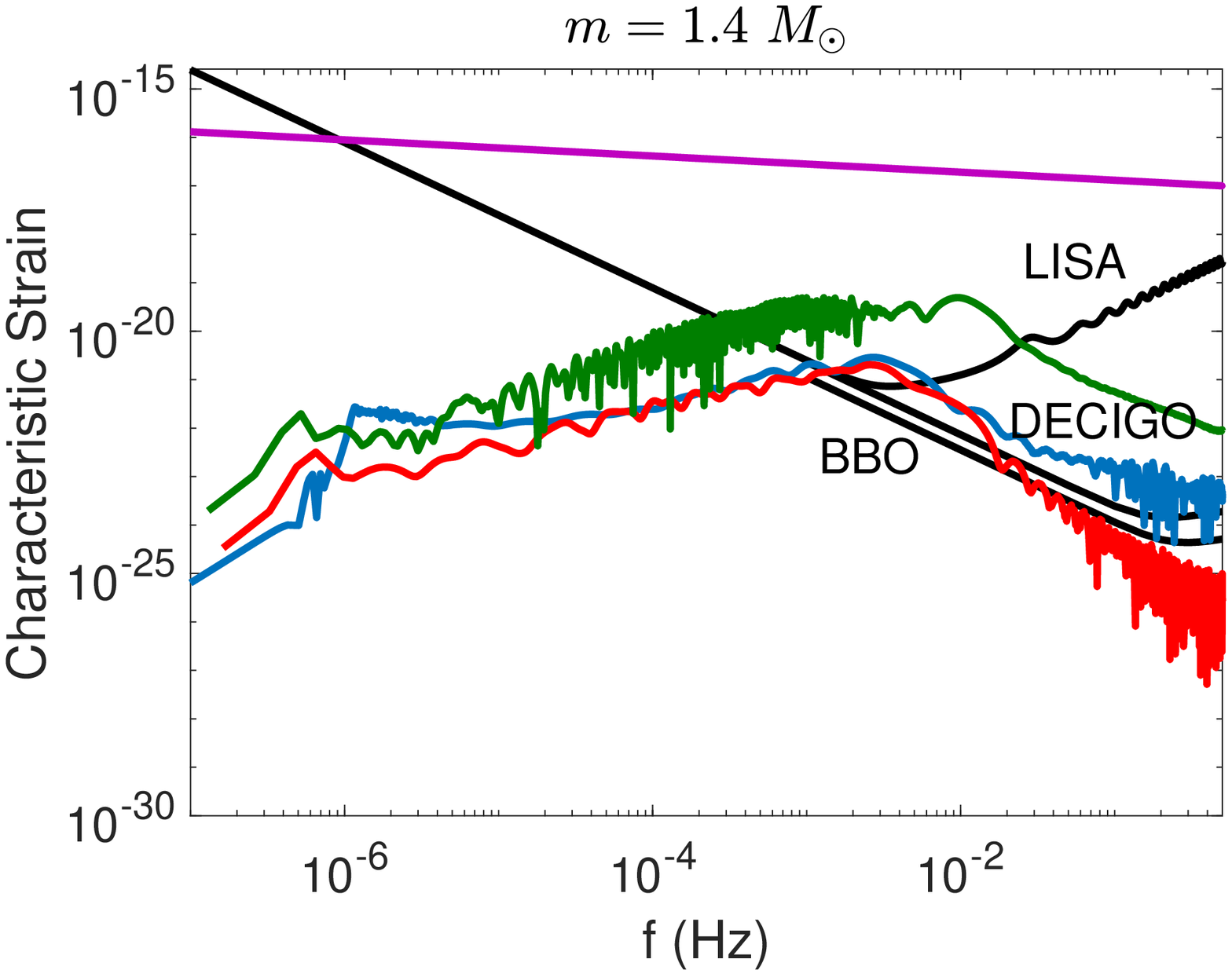}

\includegraphics[width=0.46\textwidth]{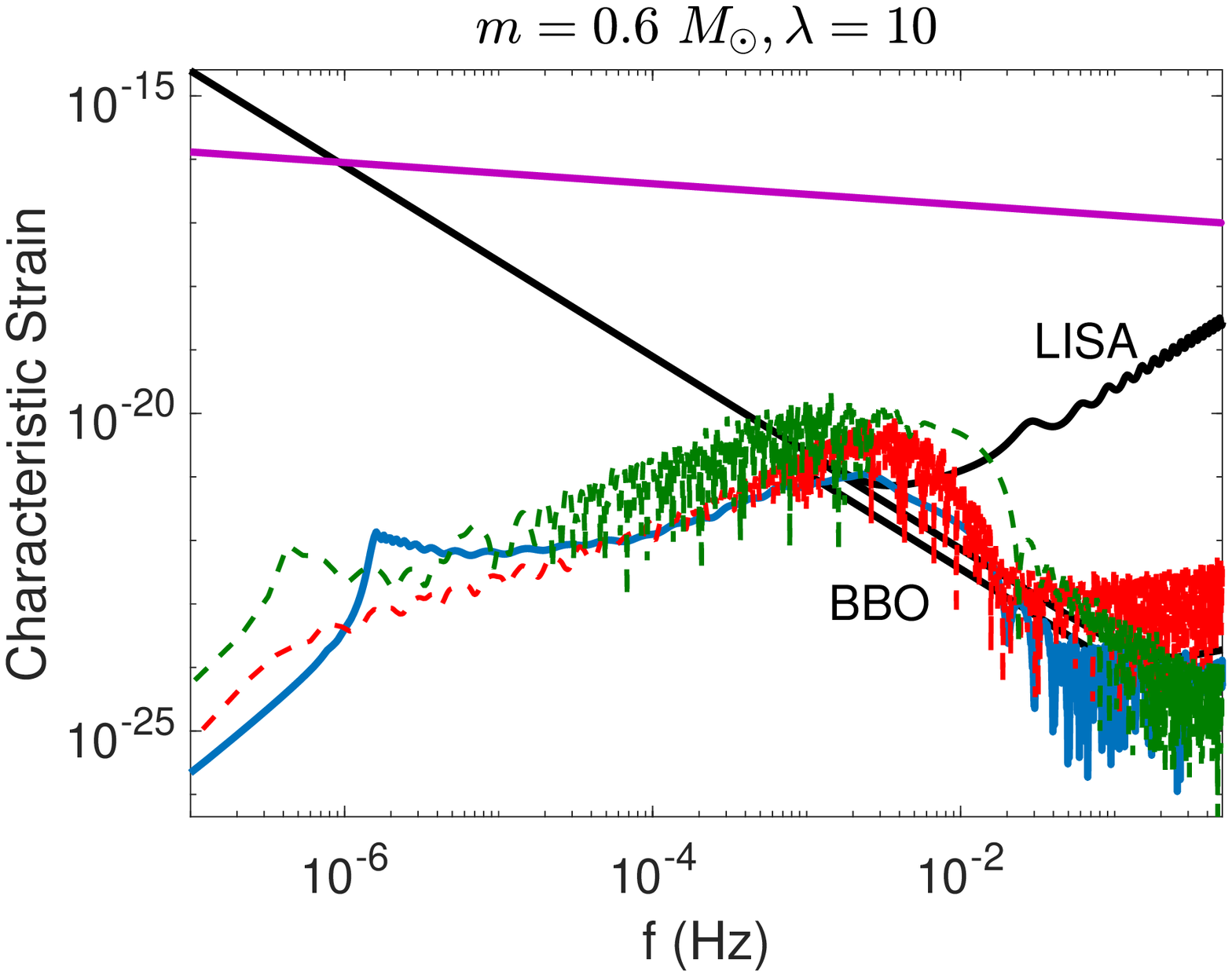}
\includegraphics[width=0.46\textwidth]{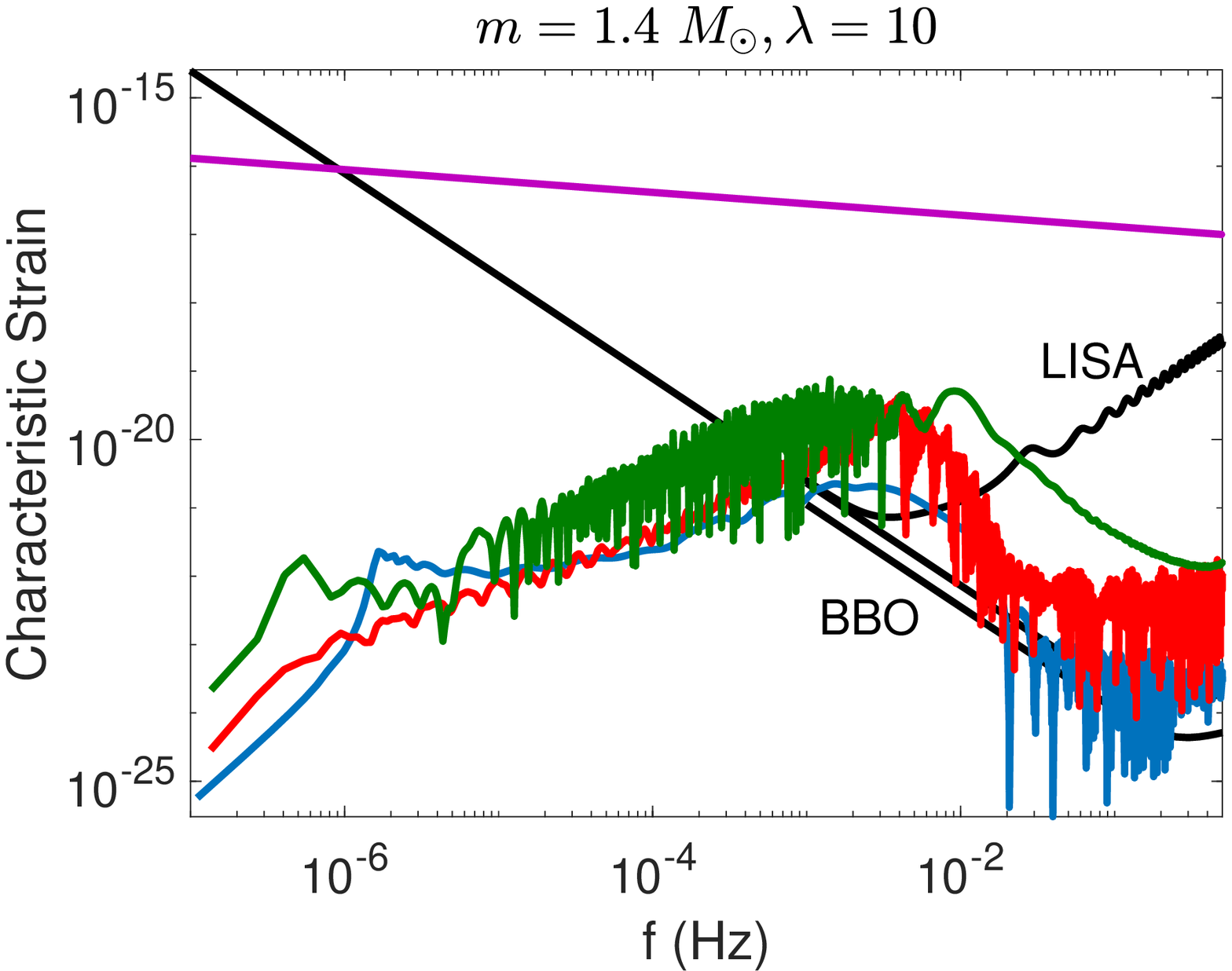}
\caption{The characteristic strain of the in-spirals
considered in this paper, together with the LISA, DECIGO and BBO sensitivity
curves, in black. The blue curves correspond to $M_{\textrm{tot}}=15M_{\odot}$,
the red ones -- to $M_{\textrm{tot}}=8M_{\odot}$, and the green ones
-- to $M_{\textrm{tot}}=5M_{\odot}$. As a reference, the violet
curve shows a binary WD merger without any envelope,
with $m_{1}=1M_{\odot},m_{2}=0.6m_{\odot}$. The left panels show
the case of $m=0.6~M_{\odot}$, and the right -- a NS companion
of $1.4~M_{\odot}$. The top row displays envelopes that remain compact,
while the bottom row shows envelopes dilated by a factor of 10 (see text). All
quantities are calculated at a distance $D=10~\textrm{kpc}$. The
Fourier transforms are calculated by sampling the signal at a rate
of $1~\textrm{Hz}$. Sampling it at higher rates (necessary
for DECIGO and BBO) should not decrease the SNRs.}
\label{fig:characteristic_strain}
\end{figure*}

The characteristic strains of the cases we considered here are presented at the top row of figure \ref{fig:characteristic_strain}. SNRs for the three detectors: LISA, DECIGO and BBO, are presented
in table \ref{tab:snr}. Note, that contrary to a gas-free GW in-spiral, $h_c(f)$ is not proportional to a power of $f$ -- it is not a straight line in logarithmic scale.

\begin{table*}
  \centering
  \caption{\label{tab:snr} SNRs for the in-spirals discussed in the text for LISA/DECIGO/BBO. $M_\textrm{tot}$ is defined as $M+M_\textrm{env}$. Cf. figure \ref{fig:characteristic_strain}.}
  \footnotesize
  \begin{tabular}{|c|c|c|c|c|c|c|c|c|c|c|}
  \hline
  ~ & ~ & \multicolumn{3}{|c|}{$M_\textrm{tot} = 15~M_\odot$} & \multicolumn{3}{c}{$M_\textrm{tot} = 8~M_\odot$} & \multicolumn{3}{|c|}{$M_\textrm{tot} = 5~M_\odot$} \\ \hline
  Envelope & $m$ ($M_\odot$) & LISA & DECIGO & BBO & LISA & DECIGO & BBO & LISA & DECIGO & BBO \\ \hline
  Compact & 0.6 & 0.57 & 1.3 & 3.4 & 0.42 & 0.25 & 0.63 & 4.6 & 1.43 & 3 \\
  Compact & 1.4 & 1.3 & 2.83 & 7.5 & 0.88 & 0.26 & 0.53 & 20 & 94 & 230 \\ \hline
  Extended & 0.6 & 0.48 & 0.33 & 0.9 & 2.3 & 6 & 19 & 5 & 1.5 & 3.2 \\
  Extended & 1.4 & 1.1 & 1.1 & 3.2 & 11 & 20 & 63 & 21 & 77 & 200 \\
  \hline
  \end{tabular}
\end{table*}

\section{Extended envelopes}
\begin{figure*}
  \centering
  \includegraphics[width=0.46\textwidth]{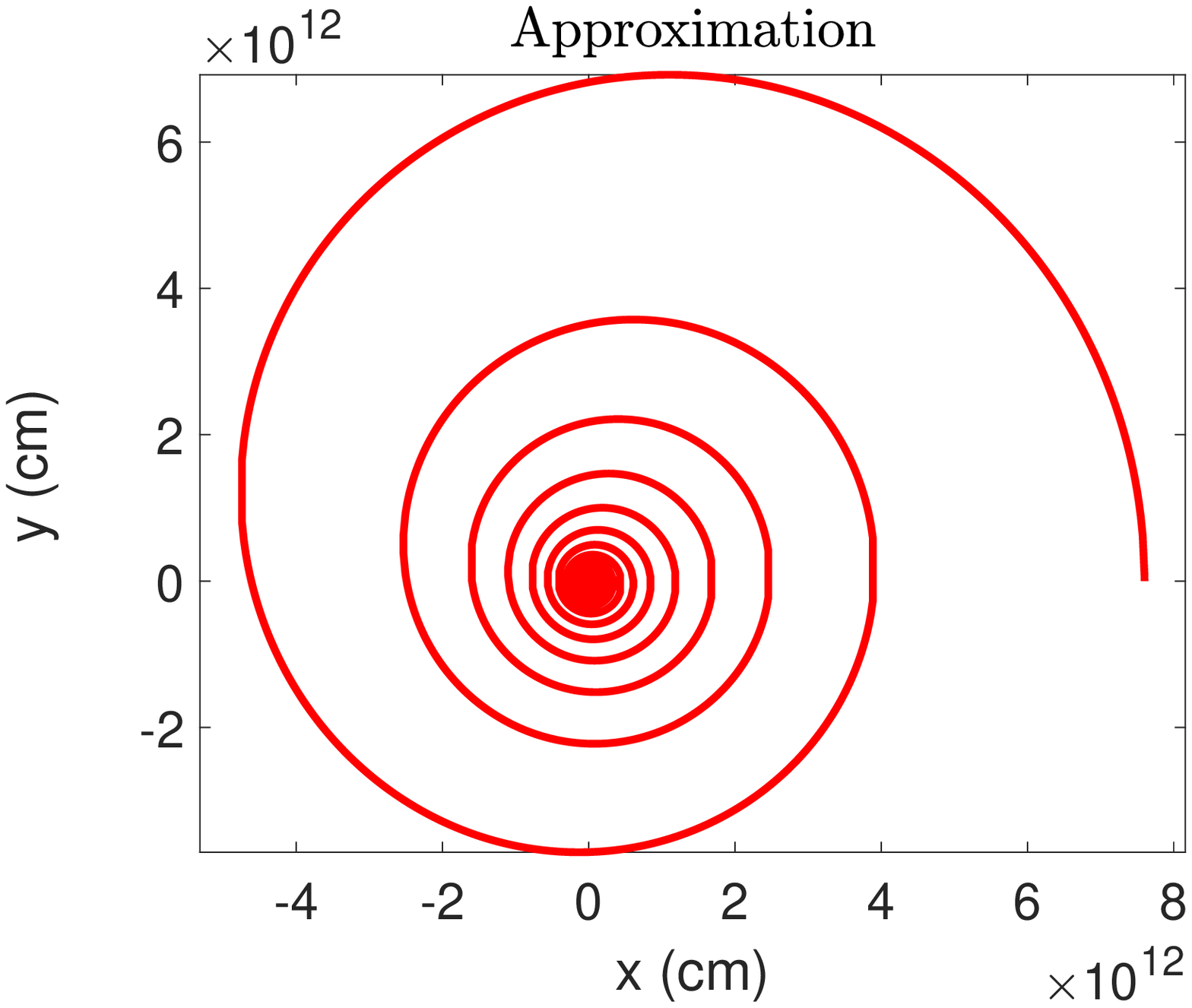}
  \includegraphics[width=0.46\textwidth]{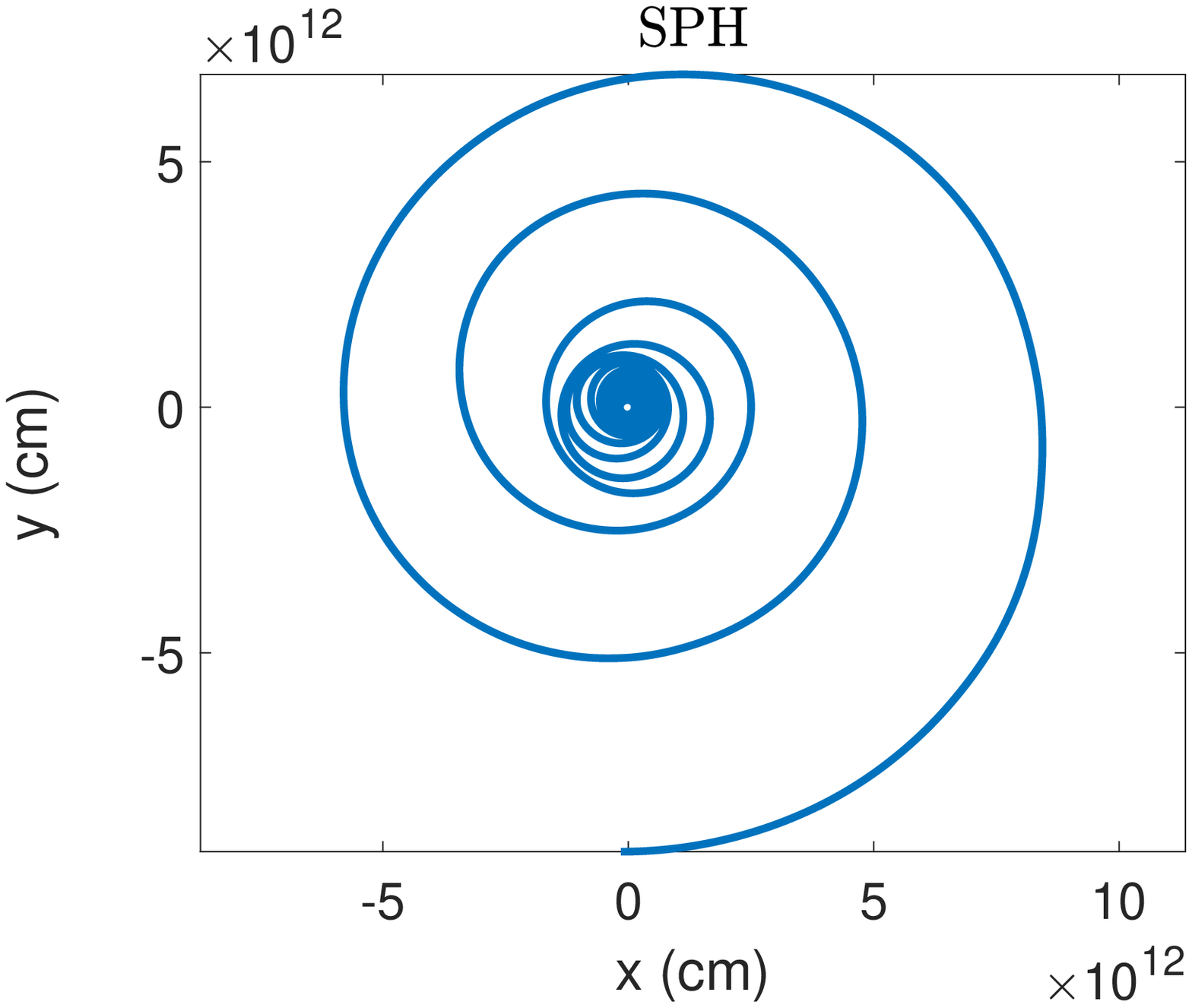}
  \caption{A comparison between the orbit -- i.e. the separation between the companion and the core -- in our approximation, taking $\lambda=11$, and in the SPH model. Both cases have $M+M_\textrm{env} = 8M_\odot, m = 0.6M_\odot$.}\label{fig:orbits comparison}
\end{figure*}

As mentioned above, the models described
here are simplified and neglect the evolution of the envelope
itself due to the in-spiral. In reality, gravitational energy is deposited into the envelope by gas dynamical friction, which inflates the envelope and results in its partial
or full unbinding \citep{Ivanova2013}. The in-spiral is therefore
expected to occur in more extended density profiles than the initial
one, and therefore the density close to
the core should be lower.
In this scenario the magnitude of the GDF force on the companion should be smaller, and the amount of time it spends
very close to the core, emitting the strongest GWs, is increased.
In other words, our SNR calculations serve as lower limits to the
expected GW signatures for such CE-in-spirals, whilst being bounded
by an upper limit -- the expected signal from a gas-free in-spiral of two COs of similar masses (see the violet line in figure \ref{fig:characteristic_strain}).
As a simplified approach we gauge the
magnitude of the extended-envelope effect using a simple, inflated
envelope model, where we again integrate equation \eqref{eqn:motion},
but we now consider a density profile which is linearly dilated, with
\begin{align}
  & \rho'(r) = \rho\left(\frac{r-R_{\textrm{core}}}{\lambda}+R_{\textrm{core}}\right)\lambda^{-3} \\ &
  c_{s}'(r)=c_{s}\left(\frac{r-R_{\textrm{core}}}{\lambda}+R_{\textrm{core}}\right)\lambda^{-1/2},
\end{align}
\emph{etc}. A plausible scale-factor given results from hydrodynamical simulations
of CE-evolution from the literature \citep{Ivanova2013} is $\lambda \approx 10$;
the characteristic strains we find for the simplified extended envelopes
are shown in the second row of figure \ref{fig:characteristic_strain}
and the total SNRs are given in table \ref{tab:snr}.
\begin{figure*}
  \centering
  \includegraphics[width=0.46\textwidth]{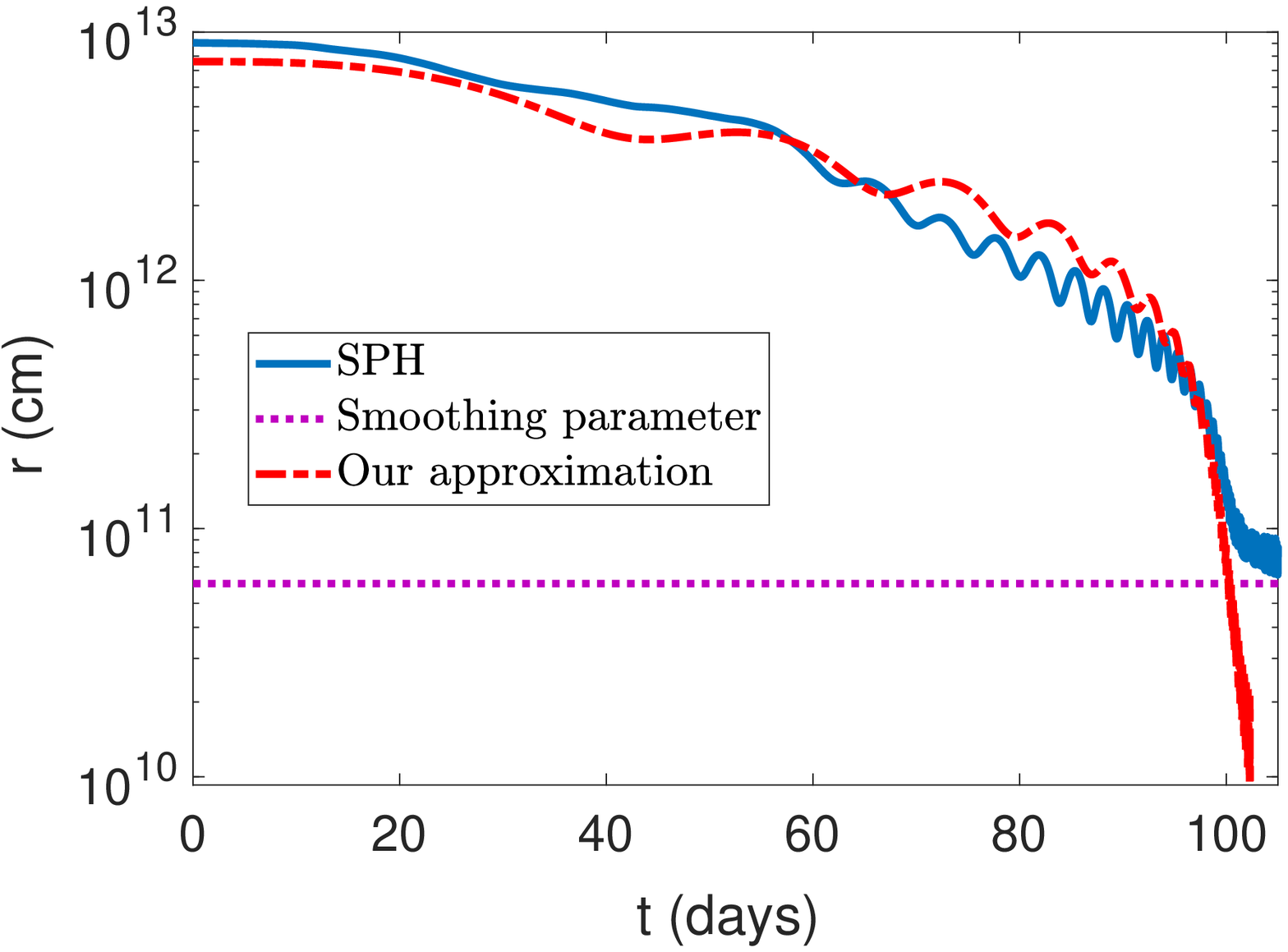}
  \includegraphics[width=0.46\textwidth]{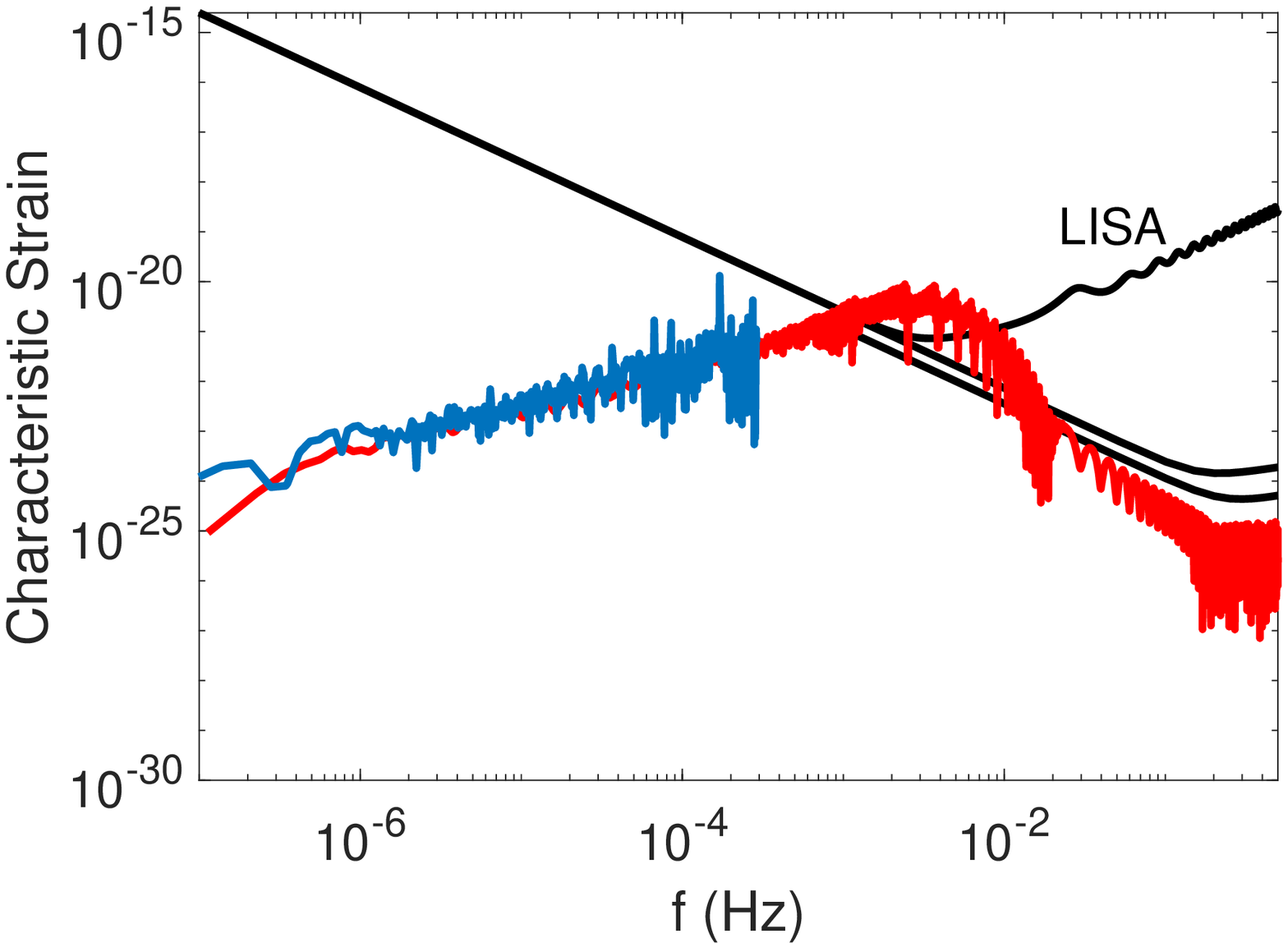}
  \caption{Left: the distances between the core and the companion, as computed from the SPH model (blue), and our approximation (red), for the case $M+M_\textrm{env} = 8M_\odot, m = 0.6M_\odot$. The smoothing parameter is shown as a horizontal dashed line. Right: the characteristic strains of the gravitational-waves emitted by the system for both cases (the SPH model was sampled once per $0.02$ days, hence its cut-off), observed from a distance of $10$ kpc. Second row: the separation between the core companion and the core in the two models (our approximation -- left; SPH -- right).}\label{fig:hydro 8-06}
\end{figure*}

As is evident from figure \ref{fig:characteristic_strain} and table
\ref{tab:snr} CE-in-spirals occurring in the Galaxy could potentially
be observed by LISA, DECIGO or BBO, where the
latter might even be sensitive to extra-Galactic sources up to
distances of a few Mpc.

To test our semi-analytical calculations, we perform a hydrodynamical
simulation for the case $M+M_\textrm{env} = 8M_\odot, m = 0.6M_\odot$. Our
simulation is run via the AMUSE framework \citep{PortegiesZwartetal2009},
which allows us to combine different type of codes. We use the same density
profile for the $8M_\odot$ red giant as above, and convert it to a
3-dimensional smoothed-particle-hydrodynamics (SPH) model with 250000 gas
particles. The core is represented by a point mass (dark-matter particle) to
avoid numerical issues and unnecessary calculation in this dense region,
whose internal changes are irrelevant for our purposes. Since the MESA model
is one-dimensional and is described by different evolution equations, we run
a relaxation phase of the SPH model alone first, for about 1000 days, while
we allow only small changes in the centre of mass velocity and position (a
restriction which is lifted later on). After we have a relaxed model, we use
it to simulate the actual CE phase with the companion which is also modelled
as a dark-matter particle. The simulation runs with the SPH code Gadget 2
\citep{Springel2005} until the core and the companion become closer than the
smoothing radius. A more thorough explanation of the simulation is presented
in \cite{Gla+18,GlanzPerets2020}.

Figure \ref{fig:hydro 8-06} displays a
comparison between the core-companion distance obtained from the SPH model,
and with our approximation, taking $\lambda = 11$. Also shown is a plot of
the characteristic strains from both cases. The comparison between our
simplified approach and the SPH model is physically meaningful only down to
the smoothing parameter, as the SPH model breaks down at lesser radii; in
this range, the two agree, thereby justifying our model even at smaller
distances which the SPH simulation cannot probe, too. Figure \ref{fig:orbits comparison} displays the orbits in the two models.

\section{Summary}
In this paper we explored gravitational-wave emission by the
in-spiral of a compact object in an evolved common-envelope binary,
leading to the final merger of the compact object with the stellar
core of the evolved star. As we show the in-spirals differ significantly from gravitational-wave in-spirals of binary isolated compact objects;
common-envelope in-spiral are dominated by a gas dynamical friction interaction
with the envelope and not by dissipation from the gravitational-wave emission
itself. Such an evolution changes the acceleration of the binary and
yields gravitational-wave signatures with a unique
frequency evolution. In particular such gravitational-wave sources could be potentially
observable by next-generation gravitational-wave detectors such as LISA/DECIGO/BBO,
and their unique frequency evolution could provide not only a smoking-gun
signature for their origin, but also enable the use of gravitational-wave detectors for mapping the interiors of evolved stars,
otherwise inaccessible through electromagnetic observations.

The common-envelope phase is also typically accompanied by explosive
electromagnetic transients. The in-spiral itself might give rise
to a luminous flare, possibly resembling stellar mergers such as V1309 Scorpii \citep{Tyl+11}. The merger of a white dwarf (or neutron star) and the degenerate core
might cause a thermonuclear explosion, potentially observable as
peculiar luminous (sub-luminous) type II/Ib supernova (given the large
amounts of hydrogen and helium in the envelopes). Finally,
the in-spiral of a neutron star into the center of the evolved star may produce
a Thorne-Żytkow object \citep{Thorne1977}. Systems with heavier compact companions, such as black holes, may generate even stronger signals. Even though there is expected to be little accretion onto such small compact objects as we consider here \citep{GlanzPerets2020}, such accretion would generate gravitational waves at frequencies that are much higher than the characteristic ones for space-based detectors \citep{Holgadoetal2018,Holgadoetal2019}, leading to a possibility of observing the same process in many detectors at the same time.

Estimates of the rates of mergers of compact objects with degenerate
cores of evolved stars leading to type Ia supernov\ae suggest that they should
occur at rates of $10-100\%$ of the observationally-inferred rates
of type Ia SNe \citep{IlkovSoker2013,Tutukovetal1992}; the rates of CEGW sources could be higher. \citet{NazinPostnov1997}
estimate a rate of $0.002~\textrm{yr}^{-1}$ for the formation of
Thorne-Żytkow objects, based on \cite{PodsiadlowskiRees1995}, implying
an overall rate of one CEGW event per a few centuries in our Galaxy. Although less likely to be observed by LISA given the expected Galactic rates, the DECIGO/BBO observatories would enable observing extra-Galactic CEGWs and potentially detect a few such events.

\section*{Acknowledgments}

We wish to thank Andei P. Igoshev and Yael Raveh for helpful discussions. YBG and EG acknowledge support from the Technion
Jacobs scholarship; and VD acknowledges support by the Israel Science Foundation (grant no. 1395/16).

\bibliographystyle{mnras}
\bibliography{gravity_df}

\begin{thebibliography}{}
\makeatletter
\relax
\def\mn@urlcharsother{\let\do\@makeother \do\$\do\&\do\#\do\^\do\_\do\%\do\~}
\def\mn@doi{\begingroup\mn@urlcharsother \@ifnextchar [ {\mn@doi@}
  {\mn@doi@[]}}
\def\mn@doi@[#1]#2{\def\@tempa{#1}\ifx\@tempa\@empty \href
  {http://dx.doi.org/#2} {doi:#2}\else \href {http://dx.doi.org/#2} {#1}\fi
  \endgroup}
\def\mn@eprint#1#2{\mn@eprint@#1:#2::\@nil}
\def\mn@eprint@arXiv#1{\href {http://arxiv.org/abs/#1} {{\tt arXiv:#1}}}
\def\mn@eprint@dblp#1{\href {http://dblp.uni-trier.de/rec/bibtex/#1.xml}
  {dblp:#1}}
\def\mn@eprint@#1:#2:#3:#4\@nil{\def\@tempa {#1}\def\@tempb {#2}\def\@tempc
  {#3}\ifx \@tempc \@empty \let \@tempc \@tempb \let \@tempb \@tempa \fi \ifx
  \@tempb \@empty \def\@tempb {arXiv}\fi \@ifundefined
  {mn@eprint@\@tempb}{\@tempb:\@tempc}{\expandafter \expandafter \csname
  mn@eprint@\@tempb\endcsname \expandafter{\@tempc}}}

\bibitem[\protect\citeauthoryear{{Abbott} et~al.,}{{Abbott}
  et~al.}{2016a}]{LIGOVirgo2016b}
{Abbott} B.~P.,  et~al., 2016a, \mn@doi [\prd] {10.1103/PhysRevD.93.122003},
  \href {https://ui.adsabs.harvard.edu/abs/2016PhRvD..93l2003A} {93, 122003}

\bibitem[\protect\citeauthoryear{{Abbott} et~al.,}{{Abbott}
  et~al.}{2016b}]{LIGOVIRGO2016}
{Abbott} B.~P.,  et~al., 2016b, \mn@doi [Physical Review Letters]
  {10.1103/PhysRevLett.116.061102}, \href
  {http://adsabs.harvard.edu/abs/2016PhRvL.116f1102A} {116, 061102}

\bibitem[\protect\citeauthoryear{{Abbott} et~al.,}{{Abbott}
  et~al.}{2018}]{LIGOVIRGO2018}
{Abbott} B.~P.,  et~al., 2018, arXiv e-prints, \href
  {http://adsabs.harvard.edu/abs/2018arXiv181112907T} {}

\bibitem[\protect\citeauthoryear{{Amaro-Seoane} et~al.,}{{Amaro-Seoane}
  et~al.}{2017}]{LISA2017}
{Amaro-Seoane} P.,  et~al., 2017, arXiv e-prints, \href
  {https://ui.adsabs.harvard.edu/\#abs/2017arXiv170200786A} {p.
  arXiv:1702.00786}

\bibitem[\protect\citeauthoryear{{Binney} \& {Tremaine}}{{Binney} \&
  {Tremaine}}{2008}]{Binney}
{Binney} J.,  {Tremaine} S.,  2008, {Galactic Dynamics: Second Edition}.
Princeton University Press

\bibitem[\protect\citeauthoryear{{Fedrow}, {Ott}, {Sperhake}, {Blackman},
  {Haas}, {Reisswig}  \& {De Felice}}{{Fedrow} et~al.}{2017}]{Fedrowetal2017}
{Fedrow} J.~M.,  {Ott} C.~D.,  {Sperhake} U.,  {Blackman} J.,  {Haas} R.,
  {Reisswig} C.,   {De Felice} A.,  2017, \mn@doi [Physical Review Letters]
  {10.1103/PhysRevLett.119.171103}, \href
  {http://adsabs.harvard.edu/abs/2017PhRvL.119q1103F} {119, 171103}

\bibitem[\protect\citeauthoryear{{Glanz} \& {Perets}}{{Glanz} \&
  {Perets}}{2018}]{Gla+18}
{Glanz} H.,  {Perets} H.~B.,  2018, \mn@doi [\mnras] {10.1093/mnrasl/sly065},
  \href {http://adsabs.harvard.edu/abs/2018MNRAS.478L..12G} {478, L12}

\bibitem[\protect\citeauthoryear{{Glanz} \& {Perets}}{{Glanz} \&
  {Perets}}{2020}]{GlanzPerets2020}
{Glanz} H.,  {Perets} H.~B.,  2020, forthcoming

\bibitem[\protect\citeauthoryear{{Grishin} \& {Perets}}{{Grishin} \&
  {Perets}}{2015}]{GP15}
{Grishin} E.,  {Perets} H.~B.,  2015, \mn@doi [\apj]
  {10.1088/0004-637X/811/1/54}, \href
  {http://adsabs.harvard.edu/abs/2015ApJ...811...54G} {811, 54}

\bibitem[\protect\citeauthoryear{{Holgado} \& {Ricker}}{{Holgado} \&
  {Ricker}}{2019}]{Holgadoetal2019}
{Holgado} A.~M.,  {Ricker} P.~M.,  2019, arXiv e-prints, \href
  {https://ui.adsabs.harvard.edu/abs/2019arXiv190210716H} {p. arXiv:1902.10716}

\bibitem[\protect\citeauthoryear{Holgado, Ricker  \& Huerta}{Holgado
  et~al.}{2018}]{Holgadoetal2018}
Holgado A.~M.,  Ricker P.~M.,   Huerta E.~A.,  2018, \mn@doi [The Astrophysical
  Journal] {10.3847/1538-4357/aab6a9}, 857, 38

\bibitem[\protect\citeauthoryear{{Ilkov} \& {Soker}}{{Ilkov} \&
  {Soker}}{2013}]{IlkovSoker2013}
{Ilkov} M.,  {Soker} N.,  2013, \mn@doi [\mnras] {10.1093/mnras/sts053}, \href
  {http://adsabs.harvard.edu/abs/2013MNRAS.428..579I} {428, 579}

\bibitem[\protect\citeauthoryear{{Ivanova} et~al.,}{{Ivanova}
  et~al.}{2013}]{Ivanova2013}
{Ivanova} N.,  et~al., 2013, \mn@doi [\aapr] {10.1007/s00159-013-0059-2}, \href
  {http://adsabs.harvard.edu/abs/2013A%26ARv..21...59I} {21, 59}

\bibitem[\protect\citeauthoryear{{Kim} \& {Kim}}{{Kim} \&
  {Kim}}{2007}]{KimKim2007}
{Kim} H.,  {Kim} W.-T.,  2007, \mn@doi [\apj] {10.1086/519302}, \href
  {https://ui.adsabs.harvard.edu/\#abs/2007ApJ...665..432K} {665, 432}

\bibitem[\protect\citeauthoryear{{Lincoln} \& {Will}}{{Lincoln} \&
  {Will}}{1990}]{LincolnWill1990}
{Lincoln} C.~W.,  {Will} C.~M.,  1990, \mn@doi [\prd]
  {10.1103/PhysRevD.42.1123}, \href
  {http://adsabs.harvard.edu/abs/1990PhRvD..42.1123L} {42, 1123}

\bibitem[\protect\citeauthoryear{Maggiore}{Maggiore}{2008}]{Maggiore}
Maggiore M.,  2008, {Gravitational Waves. Vol. 1: Theory and Experiments}.
Oxford Master Series in Physics, Oxford University Press

\bibitem[\protect\citeauthoryear{{Michaely} \& {Perets}}{{Michaely} \&
  {Perets}}{2019}]{Mic+19}
{Michaely} E.,  {Perets} H.~B.,  2019, \mn@doi [\mnras] {10.1093/mnras/stz352},
  \href {http://adsabs.harvard.edu/abs/2019MNRAS.484.4711M} {484, 4711}

\bibitem[\protect\citeauthoryear{{Moore}, {Cole}  \& {Berry}}{{Moore}
  et~al.}{2015}]{Mooreetal2015}
{Moore} C.~J.,  {Cole} R.~H.,   {Berry} C.~P.~L.,  2015, \mn@doi [Classical and
  Quantum Gravity] {10.1088/0264-9381/32/1/015014}, \href
  {http://adsabs.harvard.edu/abs/2015CQGra..32a5014M} {32, 015014}

\bibitem[\protect\citeauthoryear{{Nazin} \& {Postnov}}{{Nazin} \&
  {Postnov}}{1995}]{NazinPostnov1995}
{Nazin} S.~N.,  {Postnov} K.~A.,  1995, \aap, \href
  {http://adsabs.harvard.edu/abs/1995A%26A...303..789N} {303, 789}

\bibitem[\protect\citeauthoryear{{Nazin} \& {Postnov}}{{Nazin} \&
  {Postnov}}{1997}]{NazinPostnov1997}
{Nazin} S.~N.,  {Postnov} K.~A.,  1997, Astronomy Letters, \href
  {http://adsabs.harvard.edu/abs/1997AstL...23..139N} {23, 139}

\bibitem[\protect\citeauthoryear{{Ostriker}}{{Ostriker}}{1999}]{Ostriker1999}
{Ostriker} E.~C.,  1999, \mn@doi [\apj] {10.1086/306858}, \href
  {http://adsabs.harvard.edu/abs/1999ApJ...513..252O} {513, 252}

\bibitem[\protect\citeauthoryear{{Pani}}{{Pani}}{2015}]{Pani2015}
{Pani} P.,  2015, \mn@doi [\prd] {10.1103/PhysRevD.92.123530}, \href
  {http://adsabs.harvard.edu/abs/2015PhRvD..92l3530P} {92, 123530}

\bibitem[\protect\citeauthoryear{{Paxton}, {Bildsten}, {Dotter}, {Herwig},
  {Lesaffre}  \& {Timmes}}{{Paxton} et~al.}{2011}]{Paxtonetal2011}
{Paxton} B.,  {Bildsten} L.,  {Dotter} A.,  {Herwig} F.,  {Lesaffre} P.,
  {Timmes} F.,  2011, \mn@doi [\apjs] {10.1088/0067-0049/192/1/3}, \href
  {http://adsabs.harvard.edu/abs/2011ApJS..192....3P} {192, 3}

\bibitem[\protect\citeauthoryear{{Paxton} et~al.,}{{Paxton}
  et~al.}{2013}]{Paxtonetal2013}
{Paxton} B.,  et~al., 2013, \mn@doi [\apjs] {10.1088/0067-0049/208/1/4}, \href
  {http://adsabs.harvard.edu/abs/2013ApJS..208....4P} {208, 4}

\bibitem[\protect\citeauthoryear{{Paxton} et~al.,}{{Paxton}
  et~al.}{2015}]{Paxtonetal2015}
{Paxton} B.,  et~al., 2015, \mn@doi [\apjs] {10.1088/0067-0049/220/1/15}, \href
  {http://adsabs.harvard.edu/abs/2015ApJS..220...15P} {220, 15}

\bibitem[\protect\citeauthoryear{{Podsiadlowski}, {Cannon}  \&
  {Rees}}{{Podsiadlowski} et~al.}{1995}]{PodsiadlowskiRees1995}
{Podsiadlowski} P.,  {Cannon} R.~C.,   {Rees} M.~J.,  1995, \mn@doi [\mnras]
  {10.1093/mnras/274.2.485}, \href
  {http://adsabs.harvard.edu/abs/1995MNRAS.274..485P} {274, 485}

\bibitem[\protect\citeauthoryear{{Portegies Zwart} et~al.,}{{Portegies Zwart}
  et~al.}{2009}]{PortegiesZwartetal2009}
{Portegies Zwart} S.,  et~al., 2009, \mn@doi [\na]
  {10.1016/j.newast.2008.10.006}, \href
  {http://adsabs.harvard.edu/abs/2009NewA...14..369P} {14, 369}

\bibitem[\protect\citeauthoryear{{Springel}}{{Springel}}{2005}]{Springel2005}
{Springel} V.,  2005, \mn@doi [\mnras] {10.1111/j.1365-2966.2005.09655.x},
  \href {http://adsabs.harvard.edu/abs/2005MNRAS.364.1105S} {364, 1105}

\bibitem[\protect\citeauthoryear{{Thorne} \& {Żytkow}}{{Thorne} \&
  {Żytkow}}{1977}]{Thorne1977}
{Thorne} K.~S.,  {Żytkow} A.~N.,  1977, \mn@doi [\apj] {10.1086/155109}, \href
  {http://adsabs.harvard.edu/abs/1977ApJ...212..832T} {212, 832}

\bibitem[\protect\citeauthoryear{{Tutukov}, {Yungelson}  \& {Iben}}{{Tutukov}
  et~al.}{1992}]{Tutukovetal1992}
{Tutukov} A.~V.,  {Yungelson} L.~R.,   {Iben} Icko J.,  1992, \mn@doi [\apj]
  {10.1086/171005}, \href
  {https://ui.adsabs.harvard.edu/abs/1992ApJ...386..197T} {386, 197}

\bibitem[\protect\citeauthoryear{{Tylenda} et~al.,}{{Tylenda}
  et~al.}{2011}]{Tyl+11}
{Tylenda} R.,  et~al., 2011, \mn@doi [\aap] {10.1051/0004-6361/201016221},
  \href {http://adsabs.harvard.edu/abs/2011A%26A...528A.114T} {528, A114}

\makeatother
\end{thebibliography}

\bsp	
\label{lastpage}
\end{document}